\documentclass[11pt]{article}

\usepackage{acl}

\usepackage{times}
\usepackage{latexsym}
\usepackage{multirow}
\usepackage[T1]{fontenc}

\usepackage[utf8]{inputenc}

\usepackage{microtype}
\usepackage{ragged2e} 

\usepackage{inconsolata}

\usepackage{graphicx}
\usepackage{amsmath, amssymb}
\usepackage{xcolor}

\usepackage{fvextra}

\DefineVerbatimEnvironment{Trajectory}{Verbatim}{
  fontsize=\scriptsize,
  breaklines=true,
  breakanywhere=true,
  breaksymbolleft={},   
  breaksymbolright={},  
  frame=single,
  framesep=2mm
}

\usepackage{subcaption}

\usepackage{booktabs}
\usepackage{tabularx}
\usepackage{array}
\usepackage[section]{placeins}
\usepackage{enumitem}

\usepackage{makecell}
\usepackage{siunitx}
\usepackage{xcolor,colortbl} 
\title{ProtoCycle: Reflective Tool-Augmented Planning for Text-Guided Protein Design}

\author{
\textbf{Yutang Ge}$^{1,2}$\thanks{Equal contribution.} \quad
\textbf{Guojiang Zhao}$^{2}$\footnotemark[1] \quad
\textbf{Sihang Li}$^{3}$ \quad
\textbf{Zheng Cheng}$^{4}$ \quad
\textbf{Zifeng Zhao}$^{4}$ \\
\textbf{Hanchen Xia}$^{1}$ \quad
\textbf{Guolin Ke}$^{2}$ \quad
\textbf{Linfeng Zhang}$^{2}$ \quad
\textbf{Zhifeng Gao}$^{2}$\thanks{Corresponding authors.} \quad
\textbf{Yuguang Wang}$^{1}$\footnotemark[2] \\[0.8em]
$^{1}$Shanghai Jiao Tong University, School of Mathematical Sciences \\
$^{2}$DP Technology \\
$^{3}$University of Science and Technology of China \\
$^{4}$AI for Science Institute \\
\small{\texttt{\{lhxn0627, yuguang.wang\}@sjtu.edu.cn, \{zhaogj, gaozf\}@dp.tech}}
}

\begin{document}
\maketitle
\begin{abstract}
Designing proteins that satisfy natural language functional requirements is a central goal in protein engineering. A straightforward baseline is to fine-tune generic instruction-tuned LLMs as direct text-to-sequence generators, but this is data- and compute-hungry. With limited supervision, LLMs can produce coherent plans in text yet fail to reliably realize them as sequences. This plan--execute gap motivates \textbf{ProtoCycle}, an agentic framework for protein design that uses LLMs primarily to drive a multi-round, feedback-driven decision cycle.
ProtoCycle couples an LLM planner with a lightweight tool environment 
designed to emulate the iterative workflow of human protein engineers
and uses LLM-driven reflection on tool feedback to revise plans. Trained with supervised trajectories and online reinforcement learning, ProtoCycle achieves strong language alignment while maintaining competitive foldability, and ablations show that reflection substantially improves sequence quality. Code and resources are available at \url{https://github.com/huggggoooooo/ProtoCycle}.
\end{abstract}

\section{Introduction}
Designing proteins that achieve specified functional goals remains a central challenge with broad impact on enzyme engineering, therapeutics, and synthetic biology~\cite{kortemme2024novo,arnold2017directed,ebrahimi2023engineering}. Recent deep generative models have already shown encouraging progress toward this goal. On one hand, protein language models such as ProGen~\cite{madani_large_2023} and the ESM family~\cite{rives_biological_2021,lin_evolutionary-scale_2023} are pretrained on large sequence corpora and can be adapted to design tasks. On the other hand, text-guided design frameworks such as ProteinDT~\cite{liu_text-guided_2025}, PAAG~\cite{yuan_annotation-guided_2024}, ProDVa~\cite{liu_protein_2025}, and Pinal~\cite{Dai2024.08.01.606258} explicitly condition on natural language to generate protein sequences. 



Because requirements are expressed in natural language while proteins are discrete amino-acid sequences over a 20-letter alphabet, a straightforward baseline is to adapt a generic instruction-tuned LLM to map requirement text directly to protein sequences, following instruction-tuned generation in other domains~\cite{zhao2025molreasoner,wei2021finetuned}.
However, our analysis in Section~\ref{sec:llm-behaviour} reveals a consistent pattern: compared to direct sequence generation, LLMs appear stronger at planning in text. Under lightweight adaptation, they achieve higher plan quality and exhibit lower estimated uncertainty when producing plans, yet translating those plans into residue-level decisions that yield satisfying sequences remains challenging.

This observation aligns with how protein design is carried out in practice: a protein engineer rarely goes from a specification to an ideal sequence in a single step. Instead, reaching a satisfying design typically requires an iterative design--evaluate--revise loop over multiple rounds. Concretely, practitioners typically start from an existing scaffold, make localized edits around functional sites, and evaluate candidates using structure- or function-related signals. When the outcome is unsatisfactory, they revisit earlier choices, adjust their strategy, and iterate~\cite{wang_directed_2021,jackel_protein_2008}.

Accordingly, we propose \textbf{ProtoCycle}, a text-guided protein generation framework that casts protein design as an iterative cycle of planning, tool calling, evaluation, and revision.
In ProtoCycle, the LLM primarily serves as a high-level planner, responsible for requirement decomposition, tool selection, strategy updates, reflection, and termination decisions, while sequence generation and local editing are delegated to a lightweight set of specialized tools that can be called frequently and efficiently.
We optimise the planner in this multi-step environment with a combination of supervised fine-tuning and reinforcement learning, so that it learns when to call which tools, how to interpret feedback, and how many cycles to run before terminating.
Notably, ProtoCycle reaches competitive performance on our benchmarks using only $\sim$2{,}500 training instances for planner optimization, whereas frontier end-to-end baselines such as Pinal are trained on $\sim$1.7B protein--text pairs~\cite{Dai2024.08.01.606258}.



This work makes the following contributions:
\begin{itemize}[leftmargin=*]
    \item We empirically characterise the behaviour and limitations of generic LLMs for text-guided protein generation.
    \item We introduce a new paradigm for text-guided protein design by formulating it as a multi-round decision-making problem.
    \item We propose \textbf{ProtoCycle}, an expert-workflow-inspired agent with a lightweight toolkit.
    \item We empirically validate ProtoCycle and show via ablations that reflective revision improves sequence quality.
\end{itemize}

\section{Diagnosing Generic Instruction-tuned LLMs as Direct text-guided Protein Generators}
\label{sec:llm-behaviour}

This section studies a natural baseline: directly deploying generic instruction-tuned LLMs as zero-shot text-guided protein sequence generators.
Our goal is to test whether their strong textual reasoning reliably translates into residue-level sequences. We probe this using a simple task and a token-level uncertainty estimator~\cite{ma_estimating_2025}.

\subsection{Problem Setup}

\subsubsection{Task: text-guided Protein Generation}


We use the protein-design subset of Mol-Instructions~\cite{fang_mol-instructions:_2024} as a benchmark, where each example pairs a natural-language description with a target sequence. We run a data-scaling sanity check on Qwen2.5-7B~\cite{qwen_qwen2.5_2025} and adopt a chain-of-thought prompting format~\cite{wei_chain--thought_2023} that elicits a textual design rationale followed by an amino-acid sequence.

We study three base models (Qwen2.5-7B, Qwen2.5-72B~\cite{qwen_qwen2.5_2025}, and Llama-3.1-8B~\cite{dubey2024llama3herd}) under three usage modes: zero-shot, in-context learning~\cite{brown_language_2020}, and supervised fine-tuning (SFT).

\subsubsection{Token-level Uncertainty Estimation}

To probe model behaviour, we use Logits-induced Token Uncertainty (LogTokU)~\cite{ma_estimating_2025}, which decomposes predictive uncertainty into:
\textbf{aleatoric uncertainty (AU)}, capturing ambiguity of the next-token distribution, and
\textbf{epistemic uncertainty (EU)}, capturing how weakly supported the prediction is by the model’s evidence.
Table~\ref{tab:uncertainty-regimes} summarises the qualitative regimes for high and low AU and EU. Appendix~\ref{app:logtoku} provides a token-level illustration (Figure~\ref{fig:token-u-example}) and full formulas.

\begin{table}[t]
    \centering
    \caption{Qualitative interpretation of high/low aleatoric (AU) and epistemic (EU) uncertainty.}
    \label{tab:uncertainty-regimes}
    \small
    \begin{tabular}{lcc}
        \toprule
        & Low & High \\
        \midrule
        AU & Unambiguous problem & Ambiguous problem \\
        EU & Model is confident      & Model lacks knowledge \\
        \bottomrule
    \end{tabular}
\end{table}

\subsection{Behaviour of Generic LLMs on text-guided Protein Generation}
\label{subsec:behaviour-generic-llms}

\begin{figure*}[t]
    \centering

    \begin{subfigure}[t]{0.48\textwidth}
        \centering
        \includegraphics[
            width=\linewidth,
            height=0.15\textheight,
            keepaspectratio, trim=10 8 10 8,clip
        ]{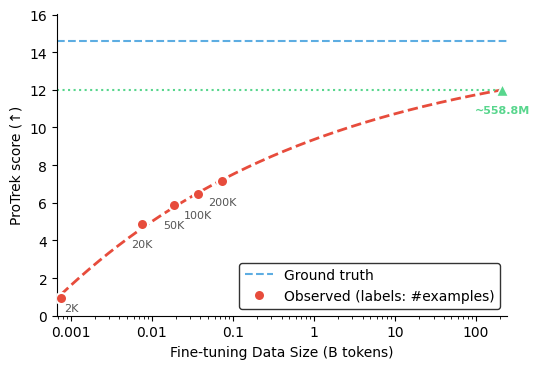}
        \caption{Effect of fine-tuning data size.}
        \label{fig:scaling-law}
    \end{subfigure}\hspace{-6mm}
    \begin{subfigure}[t]{0.48\textwidth}
        \centering
        \includegraphics[
            width=\linewidth,
            height=0.18\textheight,
            keepaspectratio
        ]{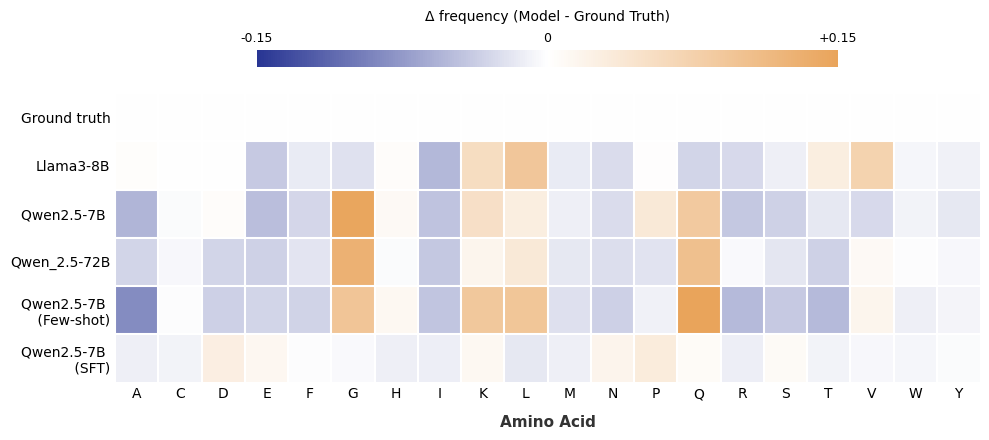}
        \caption{Deviation of amino-acid composition of generated sequences from ground truth.}
        \label{fig:aa-heatmap}
    \end{subfigure}

    \vspace{0.6em}

    \begin{subfigure}[t]{0.32\textwidth}
        \centering
        \includegraphics[
            width=\linewidth,
            height=0.16\textheight,
            keepaspectratio,
            trim=0 8 0 5,
            clip
        ]{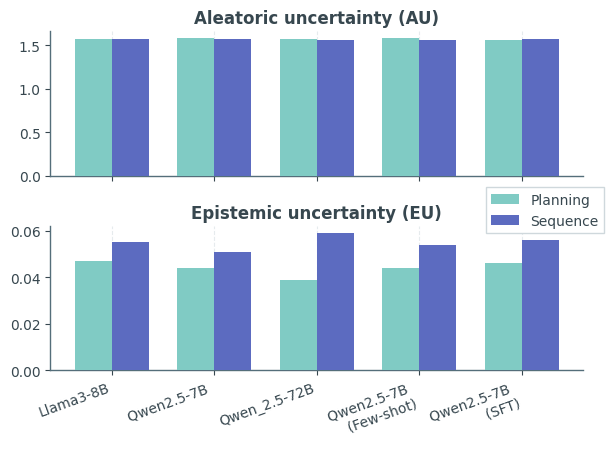}
        \caption{Aleatoric and epistemic uncertainty on planning and sequence tokens.}
        \label{fig:au-eu}
    \end{subfigure}
    \hfill
    \begin{subfigure}[t]{0.32\textwidth}
        \centering
        \includegraphics[
            width=\linewidth,
            height=0.14\textheight,
            keepaspectratio
        ]{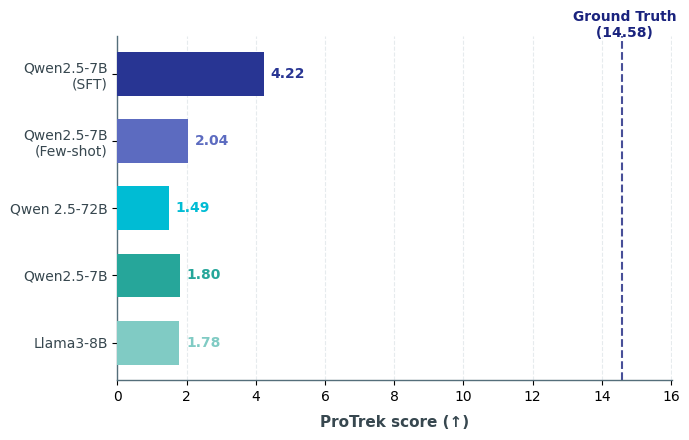}
        \caption{Text–sequence alignment score.}
        \label{fig:protrek}
    \end{subfigure}
    \hfill
    \begin{subfigure}[t]{0.32\textwidth}
        \centering
        \includegraphics[
            width=\linewidth,
            height=0.14\textheight,
            keepaspectratio
        ]{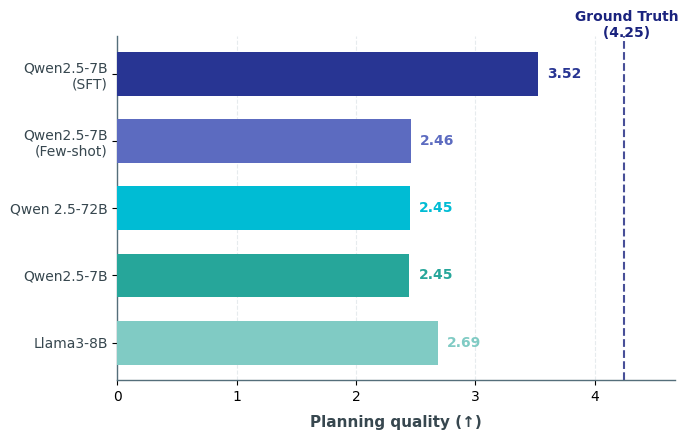}
        \caption{Quality of textual design plans.}
        \label{fig:plan-quality}
    \end{subfigure}

    \caption{
    Behaviour of generic instruction-tuned LLMs when directly used as text-guided protein generators.
    (a) Achieving strong text–sequence alignment would require very large data regimes.  
    (b) Generated amino-acid compositions exhibit clear collapse, with SFT reducing these distortions to some extend. 
    (c) Token-level uncertainty indicates that, compared to planning tokens, sequence tokens lie in a more evidence-poor regime.  
    (d) Generic LLMs lag far behind ground truth, even after SFT. 
    (e) In contrast, SFT substantially improves the quality of textual design plans, highlighting a gap between planning competence and sequence-generation ability.
    }

    \label{fig:desc2seq-behaviour}
\end{figure*}

\subsubsection{A data-scaling sanity check}

We fine-tune Qwen2.5-7B on subsets of the Mol-Instructions protein-design split (about $200$K protein-text pairs). For each run, we subsample $2$K, $20$K, $50$K, $100$K, or the full $200$K examples and fine-tune for two epochs, then evaluate text–sequence alignment on a held-out test set using the ProTrek score~\cite{su_trimodal_2025}.

Figure~\ref{fig:scaling-law} shows that more data monotonically improves ProTrek, but gains are modest: increasing the dataset from $2$K to $200$K only lifts the score from roughly $1$ to $7$, far below the ground-truth level of $\approx 14.6$. A simple power-law fit~\cite{kaplan_scaling_2020} suggests that reaching ProTrek around $12$ would require roughly $6\times 10^{8}$ supervised examples for Qwen2.5-7B, and approaching ground-truth scores requires on the order of $10^9$ pairs, similar in magnitude to Pinal~\cite{Dai2024.08.01.606258}, which trains a $16$B structure-conditioned model on $1.7$B protein–text pairs.

\subsubsection{Sequence Collapse and Language Alignment Quality}

We next analyse behaviour under the chain-of-thought protocol where the model first outputs a textual plan and then a sequence.  

We begin with marginal amino-acid usage. For each model and usage mode, we aggregate generated sequences and compare residue frequencies against those of the ground-truth sequences (see Appendix~\ref{app:aa-frequency}). Figure~\ref{fig:aa-heatmap} visualises the resulting deviations. Generic LLMs exhibit pronounced distortions: some residues are heavily overused, while others rarely appear. SFT shifts these distributions toward the ground-truth profile, alleviating the most obvious collapse patterns.

However, composition alone does not guarantee good designs. Figure~\ref{fig:protrek} reports ProTrek scores as language alignment metric. All generic LLMs remain far below ground truth, even after SFT, indicating that better marginal statistics do not translate into strong language alignment.

\subsubsection{Planning Quality}

We also evaluate the natural-language design rationales produced before sequences. Plans are scored by DeepSeek-R1~\cite{deepseek-ai_deepseek-r1:_2025} as an automatic judge; the rubric and its validation against expert ratings are given in Appendix~\ref{app:deepseek-score}.

Figure~\ref{fig:plan-quality} shows that SFT yields clear and consistent improvements in plan quality across models and usage modes. Yet the corresponding sequence-level gains are small, suggesting that the models learn to describe what should be designed much better than they learn to implement these plans at the sequence level.

\subsubsection{Uncertainty Asymmetry}

Figure~\ref{fig:au-eu} compares AU and EU on planning versus sequence tokens. AU is broadly similar in both regions, suggesting that, locally, predicting the next English word and the next residue are treated as similarly ambiguous. The asymmetry appears in EU: it is systematically higher on sequence tokens and aligns with the poor sequence quality, indicating that models operate in an evidence-poor regime when mapping requirements to concrete amino-acid choices.

Under our text-only SFT recipe, this pattern largely persists. SFT improves planning quality, but sequence-level structure and EU show no consistent gains. This supports a view in which SFT mainly refines a text-rich planning subspace, while providing too little new evidence to close the knowledge gap in the description-to-sequence mapping, so sequences become more format-plausible but remain functionally weak.

\section{Method}

\begin{figure*}[!t]
  \centering
  \includegraphics[width=0.8\linewidth]{figs/section3/picture.jpg}
  \caption{Overview of ProtoCycle compared to a human protein engineer.
  \textbf{Top}: Human workflow, which iterates between scaffold selection,
  functional-site design, evaluation, and reflection until a satisfactory
  protein is obtained. 
  \textbf{Bottom}: ProtoCycle, where a planner interacts with three tools
  (scaffold generation, functional-site design, and evaluation) via
  \texttt{<think>}/\texttt{<plan>}/\texttt{<tool\_call>} steps and receives
  summarized tool feedback to revise its strategy.}
  \label{fig:overview}
\end{figure*}

We now introduce \textbf{ProtoCycle}, which treats text-guided protein design as a multi-step decision process. 
In this view, LLM acts as a high-level planner that only decides how to design, while the actual sequence generation is delegated to a set of external tools~\cite{yao2022react,schick2023toolformer}. We first present an overview in Fig~\ref{fig:overview}, then detail the planner interface and reflection mechanism. 

\subsection{Problem Formulation}
\label{sec:prob-formulation}

Formally, given a design requirement $r$, the agent interacts with a tool environment $T$ over multiple rounds. 
At round $t$, the planner outputs decision $s_t$ and action $a_t$:
\begin{equation}
\label{eq:prob_formulation}
    (a_{t}, s_{t}) = f(s_{t-1}, t_{t-1}, \dots, s_1, t_1, r),
\end{equation}
where $t_i = T(a_i)$ denotes the tool feedback, which is a summary of tool results (see Section~\ref{sec:model-arch} for detail). 

Each action $a_i$ is factorised as
\[
a_i = (a_i^{\text{type}}, a_i^{\text{arg}}),
\]
where $a_i^{\text{type}}$ specifies which tool to invoke, and $a_i^{\text{arg}}$ specifies the input arguments for that tool.

\subsection{Tool Environment}
\label{sec:tool-env}
To emulate the expert design process under realistic compute constraints, we instantiate a lightweight tool environment $T$ with three corresponding tools: scaffold generation, functional-site design, and evaluation. Additional implementation details, query templates, and back-end mappings are provided in Appendix~\ref{app:tool-env-details}.

\paragraph{Scaffold generation}
Given a free-form requirement \texttt{text}, the scaffold generation tool
retrieves and merges candidates from multiple protein knowledge bases~\cite{uniprot2025uniprot,bansal2022rhea,blum2025interpro,binns2009quickgo}, returning $n$ scaffolds. Appendix~\ref{app:sacffolf-generation} shows more details.

\paragraph{Functional-site design.}
Given a selected \texttt{scaffold} and a local description \texttt{text}, the functional-site design tool
generates site-level variants on top of ESM2-3B~\cite{lin_evolutionary-scale_2023}. Full procedure is shown in Appendix~\ref{app:fsd-details}.

\paragraph{Evaluation.}
We evaluate candidates along two axes: \textbf{Language Alignment} and \textbf{Foldability}. Language Alignment is measured by ProTrek-650M~\cite{su_trimodal_2025}. Foldability is assessed using Chai-1~\cite{Chai-1-Technical-Report} (no-MSA), reporting pTM, pLDDT, and PAE~\cite{jumper_highly_2021}, see Appendix~\ref{app:foldability} for definitions.

\subsection{Model Architecture}
\label{sec:model-arch}
\begin{figure}[t]
  \centering
  \includegraphics[width=\linewidth]{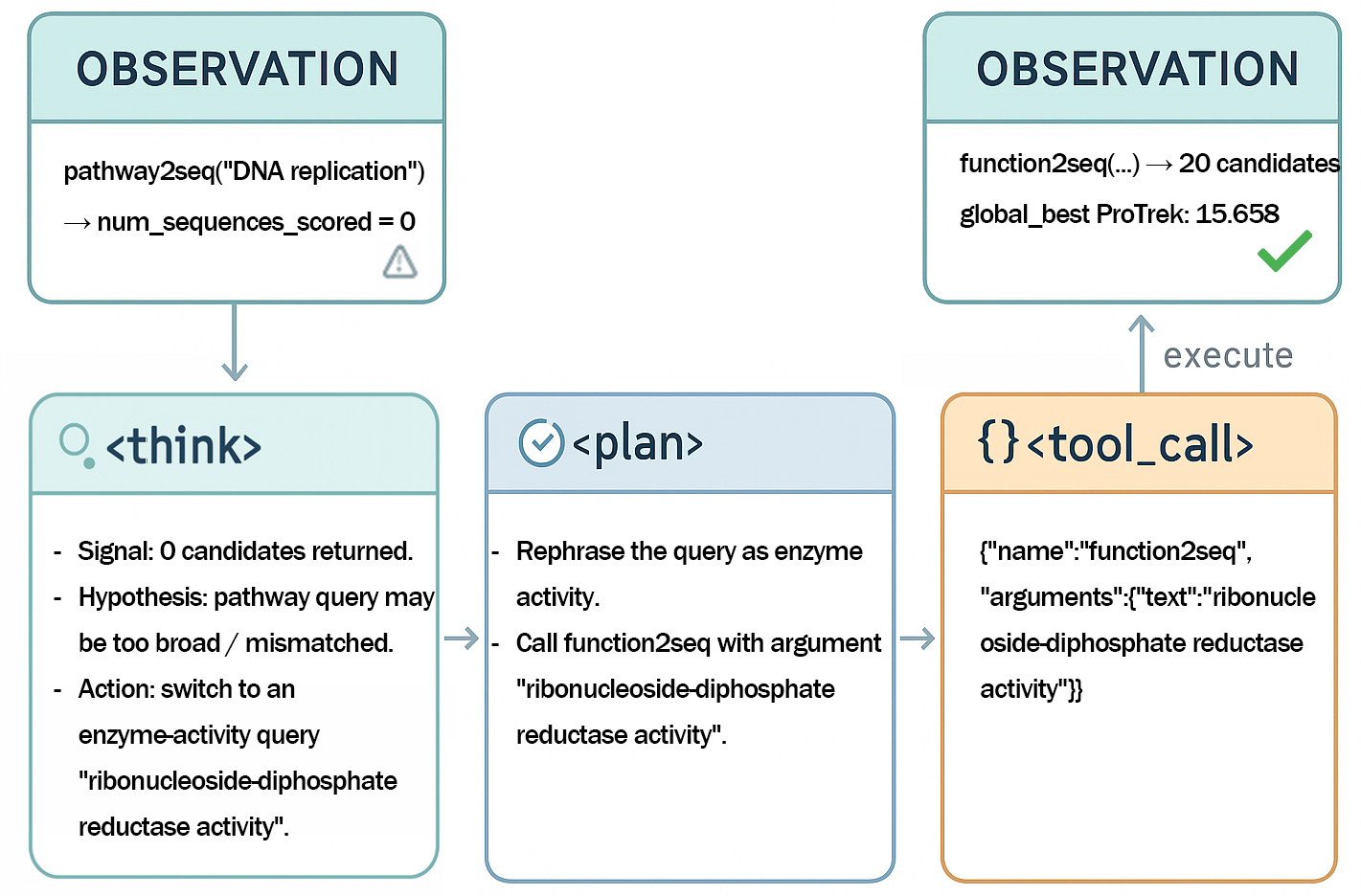}
  \caption{An example of \textbf{reflection} in ProtoCycle.}

  \label{fig:think_plan_tool}
\end{figure}

Given the output $(s_t, a_t)$ at each step as defined in~\ref{eq:prob_formulation}, the state $s_t$ consists of two parts, \texttt{<think>} and \texttt{<plan>}, while the action $a_t$ corresponds to a single \texttt{<tool\_call>} token.

\paragraph{At $t=1$ (first round).}
\begin{itemize}[leftmargin=*]
    \item \RaggedRight \texttt{<think>}: the planner decomposes the requirement $r$ into sub-requirements $\{r_1, r_2, \dots, r_n\}$, aiming to expose finer-grained design goals.
    \item \RaggedRight \texttt{<plan>}: the planner maps each $r_j$ to a tool configuration $r_j \mapsto a_j=(\text{type}(r_j), \text{arg}(r_j))$, and proposes an execution order $a_{j_1} \rightarrow a_{j_2} \rightarrow \dots \rightarrow a_{j_n}$.
\end{itemize}

\paragraph{At $t \geq 2$ (subsequent rounds).}
\begin{itemize}[leftmargin=*]
    \item \RaggedRight \texttt{<think>}: the planner summarizes the performance of the current strategy and reflects on the previous trajectory $\{s_{t-1}, t_{t-1}, \dots, s_1, t_1\}$, deciding whether to (i) continue the current plan, (ii) modify it, or (iii) terminate.
    \item \RaggedRight \texttt{<plan>}: the planner makes this high-level choice explicit, e.g., by executing the next scheduled sub-goal $r_{j_t}$ or switching to an updated sub-goal $r^{*}_{j_t}$.
\end{itemize}

After generating $s_t$ (i.e., \texttt{<think>} and \texttt{<plan>}), the planner produces the tool action $a_t = (a_t^{\text{type}}, a_t^{\text{arg}})$. In practice, we serialize $a_t$ into a JSON-like tag such as
\texttt{<tool\_call>\{ "name": $a_t^{\text{type}}$, "argument": $a_t^{\text{arg}}$ \}</tool\_call>}. A lightweight runtime then parses this tag and invokes the corresponding tool. Figure~\ref{fig:think_plan_tool} illustrates a concrete example at $t \ge 2$.
We provide a full example trajectory in Appendix~\ref{app:traj_examples}, and a complete case study in Appendix~\ref{app:case_study}.

\paragraph{Tool feedback summary.}
At each round of scaffold generation and functional-site design, the tools produce a set of sequences $P_i = \{p_{i1}, \dots, p_{im}\}$, which are scored by ProTrek-35M to obtain $U_i = \{u_{i1}, \dots, u_{im}\}$. We compute (i) the number of sequences $m$; (ii) the current-round best score $w_i=\max(U_i)$; (iii) the global best score $b_{i-1}=\max\!\Bigl(\bigcup_{j=1}^{i-1} U_j\Bigr)$; and (iv) the improvement $\Delta_i = w_i - b_{i-1}$.

The tool feedback summary $t_i$ encapsulates these statistics and serves as a compact signal of the quality of $P_i$, which the planner then uses to decide how to proceed in the next round.

\paragraph{Termination mechanism.}
When the planner decides to stop further tool use and prepare a final sequence, it triggers the evaluation tool to re-score the current top-$k$ candidates (we use $k{=}5$). The evaluation results are returned to the planner, which then either terminates and outputs the best-scoring sequence or resumes planning and invokes additional tool calls to further refine the design when improvements are still needed.

\subsection{Training}


Following recent cold-start + reinforcement learning (RL) recipes~\cite{guo2025deepseek_r1_reason,zhao2025molreasoner,li2025molr1,wei2025advancing}, we train ProtoCycle in two stages: supervised fine-tuning (SFT) to learn the basic \texttt{<think>}/\texttt{<plan>}/\texttt{<tool\_call>} protocol and tool-usage conventions, followed by online RL in the tool environment of Section~\ref{sec:tool-env}. Our training pipeline is implemented with the OpenAgentRL framework~\citep{yu2025openagentrl}.

\subsubsection{Supervised Fine-tuning}

We first train the planner with supervised fine-tuning on trajectories collected in the tool environment. Each training example is a sequence
\[
  (r, s_1, t_1, s_2, t_2, \dots, s_n, t_n).
\]

Formally, the SFT objective for one trajectory is
\[
  \mathcal{L}_{\text{SFT}}(r, s_{1:n}, t_{1:n})
  = - \frac{1}{n} \sum_{i=1}^{n}
      \log p_\theta\bigl(s_i \mid r, s_{<i}, t_{<i}\bigr),
\]
which corresponds to a standard cross-entropy loss applied only on the planner states $s_1,\dots,s_n$.

\subsubsection{Online Reinforcement Learning}
\label{sec:online-rl}

After SFT, we further optimise the planner with online reinforcement learning in the real tool environment. We treat the planner as a stochastic policy and update it with Group Relative Policy Optimization (GRPO)~\cite{shao_deepseekmath:_2024}. 

The reward $R(\tau)$ is a shaped signal that combines several components, encouraging (i) well-formed planner outputs, (ii) reasonable tool usage, (iii) reflection after poor intermediate feedback, and (iv) solving the task in a moderate number of rounds. The exact form and coefficients of these terms are given in Appendix~\ref{app:reward-details}.

\section{Experiments}

\subsection{Design from textual descriptions}
\label{sec:main-results}

\paragraph{Dataset}



We use the protein-design subset of Mol-Instructions~\cite{fang_mol-instructions:_2024}. We evaluate on three disjoint held-out test splits of 100 instances each (Eval-A, Eval-B, Eval-C) sampled from the dataset. We report main results averaged over the three splits, and provide split-wise results in Appendix~\ref{app:ci_eval}.
For SFT, we train on a disjoint set of 2{,}000 instances (see Appendix~\ref{app:sft_data} for details). For RL, we train for 5 epochs with 100 episodes per epoch, sampling requirements uniformly from the non-test pool in the real tool environment.
Training details are in Appendix~\ref{app:training_details}.

\paragraph{Baselines}
We use Qwen2.5-7B as the base language model for our planner and refer to the full agent (SFT + RL) as ProtoCycle. As generic LLM baselines in the same tool environment (Section~\ref{sec:tool-env}), we instantiate tool-interacting agents whose planners are Qwen2.5-7B, Qwen3-8B, and Qwen2.5-72B~\cite{qwen_qwen2.5_2025}, an example of the full prompting format is provided in Appendix~\ref{app:example-prompt}.
We further compare against two text-guided protein design methods: Pinal~\cite{Dai2024.08.01.606258}, a large two-stage structure-conditioned model trained on about 1.7B protein–text pairs, and ProDVa~\cite{liu_protein_2025}, which couples a text encoder and a protein LM with fragment retrieval. To isolate the effect of RL, we report both the supervised-only planner (ProtoCycle-SFT) and the planner after SFT followed by online RL (ProtoCycle-RL).

\paragraph{Metrics}
We evaluate models using three groups of metrics (formal definitions in Appendix~\ref{app:metrics}), largely following prior work on text-guided protein design~\cite{kuang2025pdfbench,liu_protein_2025}. 
\textbf{Sequence plausibility} is measured by perplexity (PPL)~\cite{jelinek1977perplexity} under ESM2-3B~\cite{lin_evolutionary-scale_2023} and the percentage of repeated residues (Repeat) following PDFBench~\cite{kuang2025pdfbench}
\textbf{Foldability} is assessed using predicted TM-score (pTM), mean per-residue confidence (pLDDT), and mean predicted aligned error (PAE)~\cite{jumper_highly_2021} from Chai-1~\cite{Chai-1-Technical-Report}, summarizing how likely a sequence is to fold into a stable three-dimensional structure. 
\textbf{Language alignment} is evaluated with the ProTrek score~\cite{su_trimodal_2025}, the EvoLLaMA score~\cite{liu2024evollama}, and retrieval accuracy~\cite{kuang2025pdfbench}, which quantify how well the designed sequences match their textual requirements in a joint embedding space.

\begin{table*}[t]
\centering
\small
\setlength{\tabcolsep}{5pt}
\begin{tabular}{l c c c c c c c c}
\toprule
\multirow{2}{*}{Model}
& \multicolumn{2}{c}{Sequence Plausibility}
& \multicolumn{3}{c}{Foldability}
& \multicolumn{3}{c}{Language Alignment} \\
\cmidrule(lr){2-3}\cmidrule(lr){4-6}\cmidrule(lr){7-9}
& PPL$\downarrow$ & Repeat$\downarrow$
& pTM$\uparrow$ & pLDDT$\uparrow$ & PAE$\downarrow$
& ProTrek$\uparrow$ & EvoLLaMA$\uparrow$ & Retrieval$\uparrow$ \\
\midrule
Natural              & 4.737 & 2.129 & 0.762 & 0.815 & 9.443 & 14.628 & 0.328 & 0.848 \\
\midrule
Qwen2.5-7B-Agent                  & 8.235 & 5.153 & 0.542 & 0.699 & 15.299 & 6.926 & 0.261 & 0.523 \\
Qwen2.5-72B-Agent                  & 7.414 & 5.341 & 0.618 & 0.714 & 13.343 & 8.791 & 0.267 & 0.563 \\
Qwen3-8B-Agent                     & 7.227 & 3.795 & 0.650 & 0.723 & 13.493 & 8.705 & 0.277 & 0.573 \\
\midrule
ProDVa                 & 5.265  & \textbf{1.580}  & 0.765 & 0.800 & 8.761  & 12.037  & 0.317 & 0.730 \\
Pinal                       & \underline{3.990}  & 9.317  & \textbf{0.792} & \textbf{0.825} & \textbf{7.768}  & \underline{14.162} & \underline{0.318} & \underline{0.807} \\
\midrule
ProtoCycle-SFT (ours)       & 4.149  & 2.902  & 0.734 & 0.807 & 10.200  & 12.502 & 0.317 & 0.840 \\
ProtoCycle-RL (ours)        & \textbf{3.865}  & \underline{2.549}  & \underline{0.775} & \underline{0.822} & \underline{8.543}  & \textbf{14.681} & \textbf{0.323} & \textbf{0.936} \\

\bottomrule
\end{tabular}
\caption{Mol-Instructions protein design results avg. over three test splits (\textbf{best}, \underline{second-best}).}
\label{tab:main-results}
\end{table*}

\paragraph{Results}

Table~\ref{tab:main-results} compares generic LLM planners, prior text-guided protein design methods, and our ProtoCycle variants on sequence plausibility, foldability, and language alignment. Our key observations are as follows.
(1) \textbf{ProtoCycle-RL is competitive with or better than specialized text-guided baselines.} It achieves the strongest language alignment overall, improving ProTrek by 3.66\% over Pinal and by 21.97\% over ProDVa, while maintaining competitive foldability: compared to Pinal, pTM/pLDDT drop by 2.15\%/0.36\% and PAE increases by 9.98\%, while compared to ProDVa we improve all foldability metrics.
(2) \textbf{Online RL yeilds substantial improvements over SFT, especially on language alignment.} Compared to ProtoCycle-SFT, ProtoCycle-RL improves ProTrek by about 17.43\% and increases retrieval accuracy by 11.43\%. 
 Meanwhile, RL also improves plausibility, reducing PPL and Repeat by 6.85\% and 12.16\%, respectively.

\subsection{Generalization to CAMEO}
\label{sec:generalization-cameo}

\begin{table}[t]
\centering
\small
\setlength{\tabcolsep}{3pt}
\renewcommand{\arraystretch}{1.05}
\begin{tabular}{lcccc}
\toprule
Model & Train & pLDDT$\uparrow$ & ProTrek$\uparrow$ & Kw. Rec.$\uparrow$ \\
\midrule
Natural      & -- & 0.79 & 10.02 & 1.0 \\
\midrule
Pinal             & $\checkmark$  & 0.75 & \textbf{11.78} & 0.39 \\
ProDVa(Mol-Inst)            & $\boldsymbol{\times}$  & 0.78 & 4.65  & 0.17 \\
ProDVa(CAMEO)          & $\checkmark$  & \textbf{0.82} & 11.05  & 0.36 \\
ProtoCycle-SFT    & $\boldsymbol{\times}$  & 0.78 & 9.08  & \underline{0.48} \\
ProtoCycle-RL     & $\boldsymbol{\times}$  & \underline{0.80} & \underline{11.17} & \textbf{0.59} \\
\bottomrule
\end{tabular}
\caption{Generalization to CAMEO (\textbf{Best}, \underline{Second Best}). Train indicates whether the model is trained on keyword-style data ($\checkmark$) or not ($\boldsymbol{\times}$). Kw. Rec. denotes keyword recovery. }
\label{tab:cameo_generalization_main}
\end{table}

\paragraph{Setup.}
We evaluate cross-dataset generalization on CAMEO~\cite{haas2018cameo}, a keyword-style protein-design dataset (see an example in Table~\ref{tab:dataset_style_comparison}).
All ProtoCycle variants use the same tool environment and evaluation protocol as in Section~\ref{sec:main-results}, and are trained only on the Mol-Instructions protein-design subset (i.e., without any keyword-style samples in the training data).
We compare against Pinal, who is trained with keyword-style data ($\checkmark$) and ProDVa with/without keyword-style training data ($\checkmark$/$\boldsymbol{\times}$). In addition to the metrics in Section~\ref{sec:main-results}, we evaluate \textbf{keyword recovery} (Kw.\ Rec.), which measures how many of the reference function keywords in the input are recovered from the designed sequence via InterProScan~\cite{jones2014interproscan}.
See Appendix~\ref{app:lang_align} for the full definition.

\paragraph{Results.}
Table~\ref{tab:cameo_generalization_main} summarizes the main metrics. Although ProtoCycle is not trained to interpret compact keyword lists as requirements, it still transfers well to CAMEO. ProtoCycle-RL achieves strong language alignment and foldability, with performance comparable to Pinal, whose train dataset contains 800M keyword-protein pairs and ProDVa (CAMEO) ,who is trained directly on the CAMEO subset with 391M training examples, demonstrating robust cross-format generalization. Full metrics are reported in Table~\ref{tab:cameo_generalization_full}.

\subsection{Tool Efficiency}
\label{sec:tool-efficiency}

\paragraph{Setup}

We quantify the effectiveness and runtime of our tool environment by invoking each tool in isolation with arguments produced by the trained ProtoCycle planner. We report standard foldability and alignment metrics on the tool outputs, and measure wall-clock latency per invocation on our implementation.

\begin{table}[t]
\centering
\footnotesize
\setlength{\tabcolsep}{2.5pt}
\renewcommand{\arraystretch}{0.9}
\resizebox{\linewidth}{!}{%
\begin{tabular}{lccc}
\toprule
Tool & ProTrek$\uparrow$ & PAE$\downarrow$ & pLDDT$\uparrow$ \\
\midrule
Scaffold search        & 11.42 & 8.96  & 0.83 \\
Functional-site design & 12.87 & 10.73 & 0.80 \\
\bottomrule
\end{tabular}}
\caption{Tool output quality when each tool is invoked once with
arguments proposed by the trained planner.}
\label{tab:tool-quality}
\end{table}

\begin{table}[t]
\centering
\small
\setlength{\tabcolsep}{4pt}
\begin{tabular}{lcc}
\toprule
Component & Latency & Output \\
\midrule
Scaffold search            & 4 s/round & $N$ retrieved sequences \\
Functional-site design     & 20 s/seq    & 1 edited sequence \\
Eval (ProTrek-35M)         & 3 s/round   & scores for $N$ seqs \\
Eval (ProTrek-650M)        & 40 s/round  & scores for $N$ seqs \\
\bottomrule
\end{tabular}
\caption{Wall-clock latency of each component.}
\label{tab:tool-runtime}
\end{table}


\paragraph{Results}
Table~\ref{tab:tool-quality} shows that, under effective prompts,
functional-site design improves language alignment relative to the raw
scaffold-search output while keeping foldability metrics broadly comparable,
suggesting that the tool can make meaningful local edits without
catastrophically degrading structural confidence. Table~\ref{tab:tool-runtime}
shows that scaffold search takes a few seconds per round (4\,s/round), whereas
functional-site design is the dominant per-sequence cost (20\,s/seq); during
interaction we therefore use ProTrek-35M for fast per-round feedback and
reserve ProTrek-650M for final reporting.

\subsection{Ablation: Reflection Mechanism}
\label{sec:ablation-reflection}

\paragraph{Setup}
To isolate the effect of reflection, we compare three planner variants under the
same tool environment and evaluation protocol as in
Section~\ref{sec:main-results}.
(i) \textbf{Workflow}: we hard-code the tool order
(\textsc{Scaffold}$\rightarrow$\textsc{Local-design}$\rightarrow$\textsc{Eval}) in the prompt.
(ii) \textbf{SFT (no reflection)}: the planner outputs \texttt{<think>} and
\texttt{<plan>} but does not perform explicit reflection on previous plans; we
fine-tune Qwen2.5-7B on non-reflective trajectories to learn the format and basic analysis.
(iii) \textbf{ProtoCycle}: our reflective planner, reported as ProtoCycle-SFT and
ProtoCycle-RL (SFT+RL).

\paragraph{Overall sequence quality}

\begin{figure}[h]
  \centering
  \includegraphics[width=0.86\linewidth]{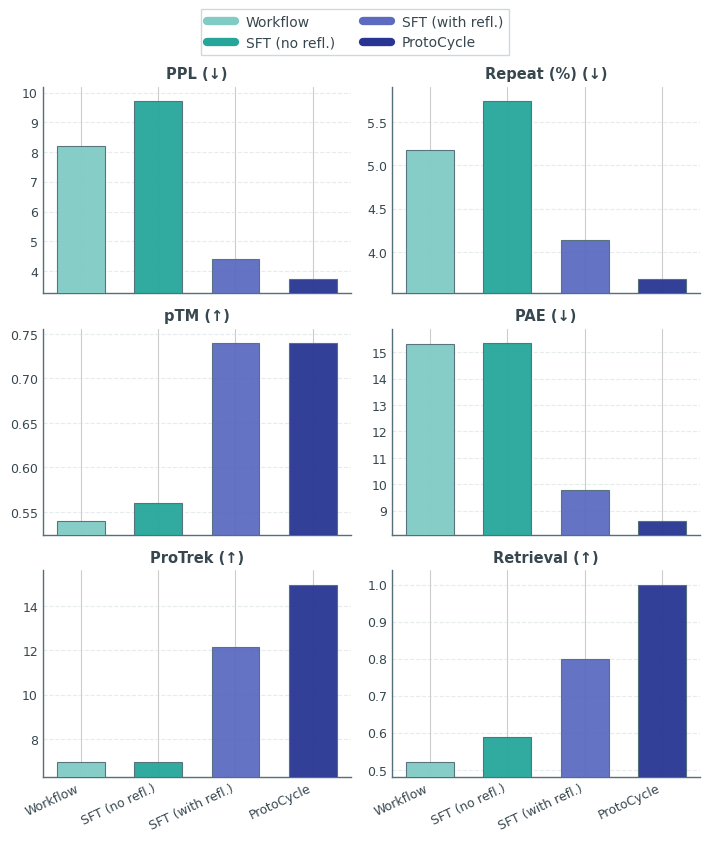}
  \caption{Reflection ablation on final plausibility, foldability, and alignment metrics.}
  \label{fig:refl-final}
\end{figure}


Figure~\ref{fig:refl-final} shows that incorporating reflection yields substantially better designs. Reflective variants achieve nearly twice the language alignment score compared to their non-reflective counterparts, and also deliver a significant improvement in foldability metrics.

Notably, SFT (no reflection) behaves similarly to the fixed
Workflow, suggesting that learning the \texttt{<think>}/\texttt{<plan>} format and
performing one-shot analysis alone is insufficient; without the ability to reflect on
tool feedback and revise strategy, the planner largely degenerates into executing a
mechanical pipeline. 

\paragraph{Effect of reflection across interaction steps}
\begin{figure}[t]
  \centering
  \begin{subfigure}[t]{0.48\linewidth}
    \centering
    \raisebox{2.5mm}{\includegraphics[width=\linewidth]{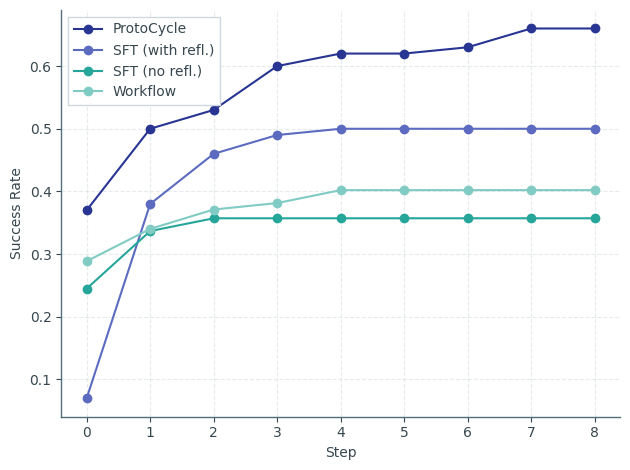}}
    \caption{Success rate vs.\ step.}
    \label{fig:refl-success}
  \end{subfigure}
  \hfill
  \begin{subfigure}[t]{0.48\linewidth}
    \centering
    \includegraphics[width=\linewidth]{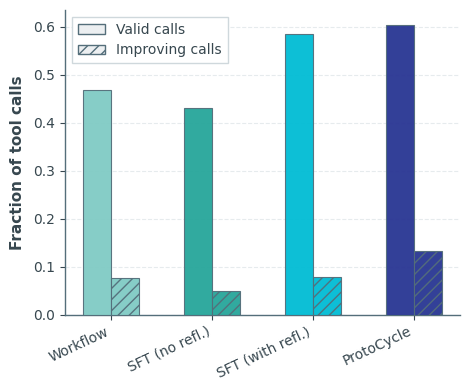}
    \caption{Valid / improving tool calls.}
    \label{fig:refl-calls}
  \end{subfigure}
  \caption{Reflection improves step-wise optimization and tool-use efficiency.}
  \label{fig:refl-step}
\end{figure}


To understand how reflection helps, we track success rate as a function of the
interaction step. For instance $n$ at step $t$, let \(\mathrm{succ}(t) = \frac{1}{m}\sum_{n=1}^{m}\mathrm{succ}_n(t),\) where \(\mathrm{succ}_n(t) = \mathbf{1}\!\left[P_n^\star(t)\ge 12\right],
P_n^\star(t) = \max\!\Bigl(\bigcup_{i=1}^{t} P_n(i)\Bigr),
\)
$P_n(i)$ is the set of ProTrek scores of all candidates produced at step $i$,
and $m$ is the number of test instances.
As shown in Figure~\ref{fig:refl-success}, reflection increases success rates by roughly $25$-$40$ \% over non-reflective baselines.
Moreover, ProtoCycle (with RL) continues to improve with more
steps, consistent with learning to explore and revise strategies based on intermediate
feedback rather than repeatedly executing a static plan.

We further quantify tool-use quality by counting (i) \textbf{valid} tool calls that
execute successfully and return non-empty outputs, and (ii) \textbf{improving} calls that
strictly increase the current best score.
Figure~\ref{fig:refl-calls} shows that reflection increases both the valid-call rate
(about $+20\%$ relative) and, more notably, the improving-call rate (about $+40\%$
relative), suggesting that reflection helps the planner not only invoke tools correctly
but also choose arguments that lead to measurable progress.

\paragraph{Reflection induces more rational decisions.}
\begin{figure}[t]
  \centering
  \begin{subfigure}[t]{0.48\linewidth}
    \centering
    \includegraphics[width=\linewidth]{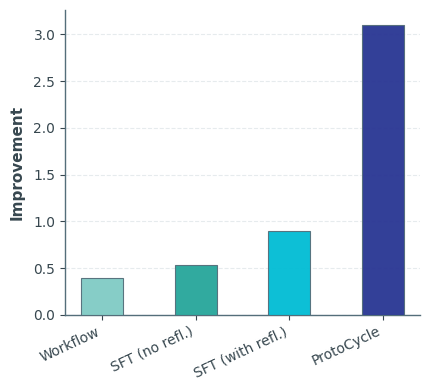}
    \caption{Net improvement over the first step.}
    \label{fig:refl-improve}
  \end{subfigure}
  \hfill
  \begin{subfigure}[t]{0.48\linewidth}
    \centering
    \includegraphics[width=\linewidth]{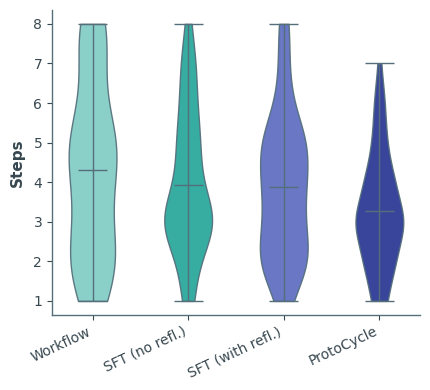}
    \caption{Wasted steps after the best score.}
    \label{fig:refl-waste}
  \end{subfigure}
  \caption{Reflection improves decision quality and compute-aware stopping.}
  \label{fig:refl-decision}
\end{figure}

Finally, we measure whether the planner makes decisions that improve upon the initial
tool outcome. For each instance $n$, we compute the net gain
$P_n^\star - P_n^\star(1)$, where $P_n^\star = \max_t P_n^\star(t)$.
Figure~\ref{fig:refl-improve} shows that without reflection the net improvement is
small, whereas reflection---especially with RL---yields substantially larger gains
(e.g., nearly a $6\times$ improvement over the Workflow baseline in our results).
This indicates that reflective planners are better at turning intermediate feedback
into concrete strategy revisions.

We also examine whether the planner stops once further tool calls become unproductive.
Let $t_{\mathrm{best}}(n)=\arg\max_t P_n^\star(t)$ and define
$\mathrm{waste}(n)=t_{\mathrm{term}}(n)-t_{\mathrm{best}}(n)$, where $t_{\mathrm{term}}$
is the terminal step.
As shown in Figure~\ref{fig:refl-waste}, reflection reduces wasted steps, indicating
that the planner can recognize diminishing returns and terminate more promptly instead
of continuing a fixed workflow.

\section{Conclusion}
\label{sec:conclusion}


Motivated by human protein-engineering workflows and our finding that generic LLMs are more reliable planners than residue-level generators, we propose ProtoCycle, which couples an LLM planner with a lightweight tool environment and an reflection mechanism for feedback-driven replanning. With only modest supervision and online interaction, ProtoCycle achieves foldability comparable to current state-of-the-art methods while outperforming them on language alignment.

\section*{Limitations}
One limitation of ProtoCycle is that our current functional-site design tool is intentionally lightweight to fit realistic compute budgets, so it can improve the odds of finding plausible candidates but cannot guarantee an "ideal" sequence that realizes the intended binding/catalytic geometry. This is consistent with broader findings in protein design: while de novo design has advanced rapidly, reliably designing complex function (especially high-specificity binding/catalysis) remains challenging and often still relies on iterative optimization and/or experimental screening rather than strict guarantees.

Another limitation is the throughput–quality trade-off inherent to an agentic workflow. Because ProtoCycle invokes structure-informed tools during planning and evaluation, it incurs higher wall-clock time and compute than one-shot generators.

\section*{Ethics Statement}
This work studies an agentic framework for text-guided protein sequence design. We do not use or collect personal user data, and our experiments rely on publicly available protein resources and benchmarks (including Mol-Instructions and CAMEO). We follow the original dataset and database licenses/terms to ensure copyright- and license-compliant use, and we apply basic quality control (e.g., validity checks and filtering) to reduce malformed or misleading samples.

Human experts were involved in a limited way during dataset construction. Specifically, experts provided a small number of seed examples that were used to prompt an external language model to synthesize additional training instances, and they performed a brief manual spot-check (around 10 samples) to flag obvious errors or inconsistencies. We did not collect or report any personally identifying information about the experts, and we treat this involvement as a lightweight sanity check rather than a comprehensive human-subject study.

We acknowledge that synthetic data and benchmark datasets can contain noise and systematic biases (e.g., uneven coverage across organisms, protein families, or functional annotations), which may propagate into model behavior and affect the diversity and reliability of generated candidates. We therefore encourage careful downstream validation and transparent reporting of failure cases when applying the method to new domains.

Finally, protein design is potentially dual-use. Our work is intended for benign scientific research and does not provide wet-lab protocols or procedural instructions for harmful applications. Any real-world deployment should follow institutional oversight and applicable biosecurity policies, including appropriate screening and usage constraints.

\bibliography{custom} 
\clearpage

\appendix

\section{Token-level Uncertainty Estimation with LogTokU}
\label{app:logtoku}

We follow the Logits-induced Token Uncertainty (LogTokU) framework~\cite{ma_estimating_2025} to obtain a token-level uncertainty score for each generated token.
At decoding step $t$, the language model produces logits $\mathbf{z}_t \in \mathbb{R}^V$ over the vocabulary.
At each decoding step the top-$K$ logits are selected,
\[
\mathcal{S}_t = \{k_1,\dots,k_K\}, \qquad
z_{t,k} = (\mathbf{z}_t)_{k}, \; k \in \mathcal{S}_t,
\]
and treated as \emph{evidence} over the small candidate set $\mathcal{S}_t$; we adopt the same procedure in our experiments.

\paragraph{Aleatoric uncertainty}
The evidence values are mapped to non-negative scalars $e_{t,k} \ge 0$ by applying a rectified linear transform to the top-$K$ logits,
\[
e_{t,k} = \max\bigl(0, z_{t,k}\bigr),
\]
and then used to define Dirichlet parameters
\[
\alpha_{t,k} = e_{t,k} + 1, \qquad
\alpha_{t,0} = \sum_{k \in \mathcal{S}_t} \alpha_{t,k}.
\]

This induces a Dirichlet distribution over the categorical probabilities of the $K$ candidates.
The \textbf{aleatoric uncertainty} (AU) at step $t$ is defined as the expected entropy of this data distribution:
\begin{equation*}
\mathrm{AU}_t
= - \sum_{k \in \mathcal{S}_t}
\frac{\alpha_{t,k}}{\alpha_{t,0}}
\Bigl[
    \psi(\alpha_{t,k}+1) - \psi(\alpha_{t,0}+1)
\Bigr],
\label{eq:logtoku-au}
\end{equation*}
where $\psi(\cdot)$ is the digamma function.
A larger $\mathrm{AU}_t$ indicates that the next-token distribution is more diffuse or multi-modal (the problem itself is more ambiguous), whereas a small $\mathrm{AU}_t$ corresponds to an essentially unambiguous next token.

\paragraph{Epistemic uncertainty}

The total amount of evidence,
\[
E_t = \sum_{k \in \mathcal{S}_t} e_{t,k},
\]
is used to quantify how much experience the model has accumulated for the current context.
The \textbf{epistemic uncertainty} (EU) at step $t$ is defined as
\begin{equation*}
\mathrm{EU}_t
= \frac{K}{E_t + K},
\label{eq:logtoku-eu}
\end{equation*}
so that $\mathrm{EU}_t$ is large when the total evidence is small (the model has little knowledge about this context), and decreases as the model accumulates more evidence.

\paragraph{Token-level uncertainty.}
The token-level predictive uncertainty is defined as:
\begin{equation}
u_t = \mathrm{EU}_t  \mathrm{AU}_t,
\label{eq:logtoku-u}
\end{equation}

Larger values of $u_t$ correspond to tokens that are both weakly supported by the model (high epistemic uncertainty) and not strongly constrained by the data (non-negligible aleatoric uncertainty), and are therefore treated as less reliable.

\section{Additional Details for LLM Experiments}
\label{app:llm-setup}

\begin{figure*}[h]
    \centering
    \includegraphics[width=\textwidth]{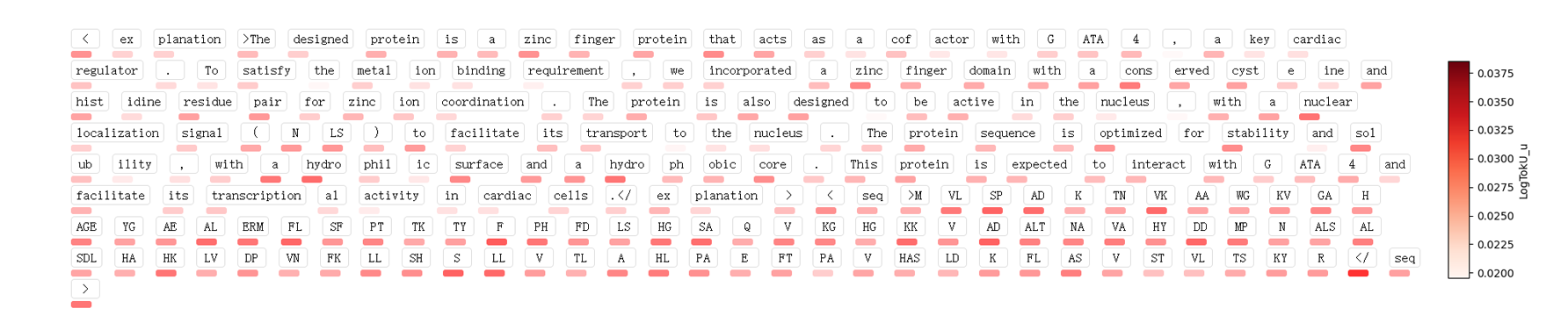}
    \caption{
    Token-level uncertainty visualization for a representative example.
    The top panel shows the generated explanation and sequence tokens, and the bottom colour bar encodes the scalar uncertainty $u_t$ (Eq.~\ref{eq:logtoku-u}) for each token.
    Regions with darker colours correspond to higher predictive uncertainty.
    }
    \label{fig:token-u-example}
\end{figure*}

\subsection{Amino-acid Frequency Estimation}
\label{app:aa-frequency}

Let the model generate sequences $s_1,\dots,s_N$, where
$s_k = (a_{k1},\dots,a_{kL_k})$ and each token $a_{kj}$ is one of the 20
standard amino acids in the alphabet $\mathcal{A}$.
For each residue type $r \in \mathcal{A}$, we estimate its empirical
frequency under the model as
\begin{equation}
    f_{\text{model}}(r)
    \;=\;
    \frac{\sum_{k=1}^N \sum_{j=1}^{L_k} \mathbf{1}[a_{kj} = r]}
         {\sum_{k=1}^N L_k},
\end{equation}
and compute the corresponding ground-truth frequency $f_{\text{gt}}(r)$
from the reference sequences in the dataset.
In Figure~\ref{fig:aa-heatmap}, we visualise the deviation
$\Delta f(r) = f_{\text{model}}(r) - f_{\text{gt}}(r)$ for each residue
type and model configuration.

\subsection{Plan-quality Scoring with DeepSeek-R1}
\label{app:deepseek-score}

We use DeepSeek-R1~\cite{deepseek-ai_deepseek-r1:_2025} as an automatic judge to evaluate
the quality of the textual design plans produced by the models.
Each plan is scored along three dimensions:
\begin{itemize}[leftmargin=*]
    \item \textbf{Framework}:
    whether the overall design strategy is coherent and logically organised;
    \item \textbf{Bio}:
    biological plausibility and correctness of the proposed operations;
    \item \textbf{Alignment}:
    how well the plan addresses the given textual requirement.
\end{itemize}
The final planning quality score is computed as a weighted combination of
these components:
\begin{equation*}
\begin{aligned}
    \text{PlanScore}
    &= 0.4 \cdot \text{framework} \\
    & + 0.3 \cdot \text{bio} \\
    & + 0.3 \cdot \text{alignment}.
\end{aligned}
\end{equation*}

To validate this automatic evaluation, we randomly select 20 descriptions
and obtain human expert ratings for both the framework and alignment of
the generated plans.
The overall DeepSeek score shows a strong correlation with human
judgements (Pearson $r = 0.73$), and the alignment sub-score correlates
with the ProTrek text--text similarity between the plan and the
requirement (Pearson $r = 0.65$).
These results suggest that DeepSeek provides a reasonably faithful proxy
for expert assessment of planning quality.

\section{Tool Environment Details}
\label{app:tool-env-details}

\subsection{Scaffold Generation}
\label{app:sacffolf-generation}
{\RaggedRight The scaffold-generation tool
\(
T(a^{\textsc{scaf}}, (\texttt{text}, \text{num\_sequences}))
\)
wraps several database-specific queries behind a unified interface. The argument \texttt{text} is first lightly normalized (lowercasing and simple keyword matching) and then routed to one or more back-ends depending on its content:

\begin{itemize}[leftmargin=*]
  \item general functional descriptions (e.g., enzyme names, cofactor phrases) $\rightarrow$ UniProtKB~\cite{uniprot2025uniprot};
  \item pathway- or reaction-related cues $\rightarrow$ Rhea~\cite{bansal2022rhea};
  \item GO terms or GO-style phrases $\rightarrow$ QuickGO~\cite{binns2009quickgo};
  \item explicit domain / motif / structural keywords $\rightarrow$ InterPro~\cite{blum2025interpro}.
\end{itemize}
}

Each back-end returns a small set of candidate proteins together with basic annotations; we extract their amino-acid sequences as candidate scaffolds and merge them into a single list. When the union exceeds \texttt{num\_sequences}, we subsample using a simple ranking heuristic that favors proteins with richer annotations and lengths within a task-specific window. The mapping between query types and back-end databases is summarized in Table~\ref{tab:scaffold-tools}.

\begin{table}[h]
    \centering
    \caption{Query types and back-end resources used by the scaffold-generation tool.}
    \label{tab:scaffold-tools}
    \small
    \begin{tabular}{ll}
        \toprule
        Query type  & Back-end database \\
        \midrule
        General functional description  & UniProtKB \\
        Pathway / reaction information & Rhea \\
        GO terms or GO-style phrases   & QuickGO \\
        Domain / structural motif cues & InterPro \\
        \bottomrule
    \end{tabular}
\end{table}

\subsection{Implementation Details of the Functional-site Design Tool}
\label{app:fsd-details}

\paragraph{Identifying and locking annotated functional residues}
Given a scaffold sequence 
\(
  \mathbf{s} = (s_1,\dots,s_L)
\),
we query UniProt for residue-level annotations across five feature types:
\texttt{binding\_site}, \texttt{active\_site}, \texttt{motif}, \texttt{domain}, and \texttt{region}.
Each annotation returns an interval \([i,j]\) on the sequence. We define the set of \emph{locked} positions as
\[
  \mathcal{L} \;=\; \bigcup_{(i,j) \in \mathcal{A}} \{i,\dots,j\},
\]
where \(\mathcal{A}\) is the union over all returned intervals. Residues in \(\mathcal{L}\) are treated as immutable: subsequent motif insertions and local edits are restricted to indices \(t \notin \mathcal{L}\). This preserves known functional or structurally critical regions while allowing the tool to operate on less constrained parts of the scaffold.

\paragraph{Motif retrieval and ESM-based likelihood scoring}
The textual specification \texttt{text} describes the desired local function (e.g., "binds Zn$^{2+}$ with a Cys$_2$His$_2$ coordination motif"). We use \texttt{text} as a query to UniProt~\cite{uniprot2025uniprot} (optionally restricted by organism or keyword filters) and collect a small set of documented motif sequences
\(
  \mathcal{M} = \{\mathbf{m}^{(1)},\dots,\mathbf{m}^{(K)}\}
\),
each with length \(|\mathbf{m}^{(k)}| = \ell_k\).

For a candidate motif \(\mathbf{m}\) and an admissible insertion window \([p, p+\ell-1]\) that does not intersect \(\mathcal{L}\), we construct a hypothetical edited sequence
\[
  \mathbf{x}^{(\mathbf{m},p)} = (x^{(\mathbf{m},p)}_1,\dots,x^{(\mathbf{m},p)}_L),
\]
by replacing positions \(p,\dots,p+\ell-1\) of \(\mathbf{s}\) with \(\mathbf{m}\) (or inserting \(\mathbf{m}\) into a loop if allowed by the backbone representation). We then score this candidate using ESM2-3B~\cite{lin_evolutionary-scale_2023}.

Because ESM2-3B is a masked language model, we estimate a \emph{pseudo} negative log-likelihood (NLL) over a local region \(\Omega^{(\mathbf{m},p)}\) around the graft (e.g., the motif plus a small flanking window):
\[
  \mathrm{NLL}\bigl(\mathbf{x}^{(\mathbf{m},p)}\bigr)
  \;=\;
  -\sum_{t \in \Omega^{(\mathbf{m},p)}} 
  \log p_{\theta}\bigl(x^{(\mathbf{m},p)}_t \,\big|\, \mathbf{x}^{(\mathbf{m},p)}_{\setminus t}\bigr),
\]
where \(p_{\theta}(\cdot \mid \mathbf{x}_{\setminus t})\) is the ESM2-3B conditional distribution obtained by masking position \(t\) and running a single forward pass.
Intuitively, this score penalizes candidates that are implausible under the protein language model, while leaving the rest of the scaffold unchanged.

Across all motifs \(\mathbf{m} \in \mathcal{M}\) and all admissible insertion windows \(p\), the tool selects the configuration
\[
  (\mathbf{m}^\star, p^\star) 
  \;=\; \arg\min_{\mathbf{m},\,p} \;
  \mathrm{NLL}\bigl(\mathbf{x}^{(\mathbf{m},p)}\bigr),
\]
and returns \(\mathbf{x}^{(\mathbf{m}^\star,p^\star)}\) as the proposed functional-site design (subject to additional filtering below).


\section{Reward Shaping Details}
\label{app:reward-details}

Given a trajectory $\tau$ with $T$ rounds, the final reward is
\begin{equation*}
  R(\tau) =
  \begin{cases}
    R_{\text{soft}}(\tau), & \text{valid \texttt{<answer>} in $\tau$}, \\[2pt]
    -0.5,                  & \text{otherwise}.
  \end{cases}
\end{equation*}

The shaped component is
\begin{align*}
  R_{\text{soft}}(\tau)
  &= 0.5\,R_{\text{format}}(\tau)
   + 1.0\,R_{\text{tools}}(\tau) \notag \\
  &\quad
   + 0.5\,R_{\text{ProTrek}}(\tau)
   + 1.5\,R_{\text{eff}}(T)
   + R_{\text{refl}}(\tau).
\end{align*}

\paragraph{Format reward}
Let $F_{\text{tp}}(\tau)$ be $1$ if $\tau$ contains at least one pair of
\texttt{<think>} and \texttt{<plan>} tags, and $0$ otherwise.  Let
$F_{\text{ans}}(\tau)$ be $1$ if $\tau$ contains a valid
\texttt{<answer>} tag.  We set
\begin{equation*}
  R_{\text{format}}(\tau)
  = 0.5\,F_{\text{tp}}(\tau)
    + 0.5\,F_{\text{ans}}(\tau).
\end{equation*}

\paragraph{Sequence-quality reward (ProTrek)}
Let $P^\star(\tau)$ be the best ProTrek score over all sequences in
$\tau$, and let $S(\tau) \in \{0,1\}$ indicate whether at least one
sequence was generated. Then
\begin{equation*}
  R_{\text{ProTrek}}(\tau) =
  \begin{cases}
    \mathrm{clip}\bigl(P^\star(\tau)/20,\,0,\,1\bigr), & S(\tau) = 1, \\[2pt]
    -0.2,                                              & S(\tau) = 0.
  \end{cases}
\end{equation*}

\paragraph{Efficiency reward}
Let $T$ be the total number of rounds. We first separate the two regimes
\begin{equation*}
  R_{\text{eff}}(T) =
  \begin{cases}
    0.1\,(4 - T), & T \le 4, \\[2pt]
    r_{\text{mid}}(T), & T > 4,
  \end{cases}
\end{equation*}
and then define
\begin{align*}
  r_{\text{mid}}(T)
  = &-0.25\,(T - 4)
    - 0.5\,\mathbf{1}[T > 8] \\
    &- 1.0\,\mathbf{1}[T > 12].
\end{align*}

\paragraph{Tool-usage reward}
For each round $i$, we define a local tool reward $r_{\text{tool}}(a_i,t_i)$
and sum over the episode:
\begin{equation*}
  R_{\text{tools}}(\tau)
  = \sum_{i=1}^{T} r_{\text{tool}}(a_i, t_i).
\end{equation*}
In our implementation, $r_{\text{tool}}(a_i,t_i)$ is
(i) a small positive constant $\alpha_{\text{gen}} = 0.3$–$0.6$
for the first few successful calls to generative tools,
(ii) an extra bonus $\alpha_{\text{score}} = 0.3$ for
successful \texttt{get\_score} calls, and
(iii) a penalty of $-0.1$ for failed tool calls.

\paragraph{Reflection reward}
Let $\mathrm{bad}(t_i) \in \{0,1\}$ indicate that the feedback at round
$i$ is below a threshold, and let $\mathrm{refl}(s_{i+1}) \in \{0,1\}$
indicate that the next state contains an explicit reflection step.  We
define
\begin{align*}
  R_{\text{refl}}(\tau)
  &= \sum_{i=1}^{T-1} r_{\text{refl}}(t_i, s_{i+1}), \\[2pt]
  r_{\text{refl}}(t_i, s_{i+1})
  &= \alpha_{\text{good}}\,
     \mathrm{bad}(t_i)\,\mathrm{refl}(s_{i+1}) \notag \\
  &\quad
     + \alpha_{\text{noop}}\,
       \mathrm{bad}(t_i)\,\bigl(1 - \mathrm{refl}(s_{i+1})\bigr),
\end{align*}
with $\alpha_{\text{good}} > 0$ and $\alpha_{\text{noop}} < 0$
(e.g.\ $\alpha_{\text{good}} = 0.2$ and $\alpha_{\text{noop}} = -0.2$).


\section{Training Details}
\label{app:training_details}

This appendix provides the detailed training setup for ProtoCycle, including
data construction, SFT settings, RL settings, tool-environment interface,
and compute resources.

\subsection{SFT Trajectory Synthesis}
\label{app:sft_data}
\subsubsection{Overview}

We construct 2{,}000 multi-round tool-interaction trajectories for SFT by seeding \textsc{GPT-4o} with 5 expert-written trajectories as few-shot demonstrations. To mimic the reflection mechanism, we additionally prompt \textsc{GPT-4o} to condition its next-step decision on whether the previous tool call succeeded and on the returned ProTrek score: (i) if the tool call fails, it either retries with revised arguments or switches to an alternative tool; (ii) if the tool call succeeds but the ProTrek score is below 12, it reflects on potential issues in the current workflow and replans; (iii) if the tool call succeeds and the ProTrek score is above 12, it continues with the current plan.

Each trajectory follows an interaction pattern with \texttt{<think>}, \texttt{<plan>}, and \texttt{<tool\_call>} segments, interleaved with tool observations. We additionally score the initial global plan of each synthesized trajectory using \textsc{DeepSeek-R1} following the rubric in Appendix~\ref{app:deepseek-score}; plans that do not meet the quality threshold are discarded and the trajectory is regenerated.

\subsubsection{Example Tool-Interaction Trajectories}
\label{app:traj_examples}


\begin{Trajectory}
[USER REQUIREMENT]
1. The protein should have a specific function in intein-mediated protein splicing, cell-cell signaling and should be localized to plasma membrane, endoplasmic reticulum membrane, Golgi membrane in order to carry out this function efficiently.
2. The designed protein have Hint domains with specific sequence motifs that are crucial for its activity.

[ASSISTANT]
<think>
[Requirement decomposition]
- general function: "specific function in intein-mediated protein splicing, cell-cell signaling".
- pathway: "intein-mediated protein splicing\", "cell-cell signaling"
- domain: "Hint domains".
- go: "plasma membrane", "endoplasmic reticulum membrane", "Golgi membrane".
- motif: "specific sequence motifs that are crucial for its activity"
- signal: localization to "plasma membrane, endoplasmic reticulum membrane, Golgi membrane
</think>
<plan>
- First, explore functional scaffolds using pathway2seq with "intein-mediated protein splicing" as the pathway keyword, since it directly relates to the protein's function.
- Once initial scaffolds are generated, introduce motif2constraints to enforce the presence of specific sequence motifs and signal2constraints to ensure proper localization.
- After constraints are applied, call get_score to rank sequences and stop when the global best score is high and stable across multiple rounds.
</plan>
<tool_call>
{"name": "pathway2seq", "arguments": {"text": "intein-mediated protein splicing"}}
</tool_call>

[TOOL OBSERVATION]
Round: 1
- tool: pathway2seq
- num_sequences_scored: 0
- delta_best: +0.000
- argument_pathway: intein-mediated protein splicing

[ASSISTANT]
<think>
The first attempt to generate scaffolds using the pathway2seq tool with the pathway "intein-mediated protein splicing" did not yield any candidate sequences. This indicates that the argument may have been too specific or not aligned with the tool's expectations. 
To proceed, I will try a different approach by using the function2seq tool with a more general description of the protein's function. The phrase "intein-mediated protein splicing, cell-cell signaling" can be simplified to "intein-mediated splicing and cell signaling" to see if it retrieves any candidate sequences.
This should help in generating a broader range of scaffolds that align with the protein's functional requirements.
</think>
<plan>
Call the function2seq tool with the argument "intein-mediated splicing and cell signaling" to generate candidate protein sequences.
</plan>
<tool_call>
{"name": "function2seq", "arguments": {"text": "intein-mediated splicing and cell signaling"}}
</tool_call>

[TOOL OBSERVATION]
Round: 2
- tool: function2seq
- num_sequences_scored: 50
- best_ProTrek_this_round: 13.858
- global_best_ProTrek: 13.858
- delta_best: +13.858

[ASSISTANT]
<think>
The function2seq tool successfully generated 20 candidate sequences, with the best ProTrek score reaching 13.858. This is a strong score, indicating that the scaffold aligns well with the functional requirements. However, the design still needs to incorporate specific constraints related to localization and sequence motifs.
To ensure proper localization to the plasma membrane, endoplasmic reticulum membrane, and Golgi membrane, I will introduce motif2constraints to enforce the presence of specific sequence motifs crucial for the protein's activity.
</think>
<plan>
Call the motif2constraints tool to enforce the presence of specific sequence motifs that are crucial for the protein's activity.
</plan>
<tool_call>
{"name": "motif2constraints", "arguments": {"motif": "specific sequence motifs that are crucial for its activity"}
</tool_call>

[TOOL OBSERVATION]
Round: 3
- tool: motif2constraints
- num_sequences_scored: 5
- global_best_ProTrek: 14.476
- prev_global_best_ProTrek: 13.858
- delta_best: +0.618
- argument_motif: specific sequence motifs that are crucial for its activity

[ASSISTANT]
<think>
The motif constraint has improved the global ProTrek score, suggesting the current top candidate is already aligned with the requirement. 
Before allocating additional tool calls, I will run the evaluator once to obtain calibrated scores for language alignment and foldability of the current best candidate.
</think>
<plan>
Call get_score to evaluate the current top candidates and verify requirement satisfaction.
</plan>
<tool_call>
{"name": "get_score", "arguments": {"top_k": 5}}
</tool_call>

[TOOL OBSERVATION]
Round: 4
- tool: get_score
- num_sequences_scored: 5
- ProTrek_best: 14.421
- pTM_best: 0.78
- pLDDT_best: 0.81
- PAE_best: 9.4

[ASSISTANT]
<think>
Given that the score is strong and stable, I will stop and return the current best candidate as the final design.
</think>
<answer>
MDE...KSS
</answer>

\end{Trajectory}

\subsubsection{Prompt for ProtoCycle}
\label{app:example-prompt}
In our implementation, tool-related instructions (i.e., the tool list, schemas, and calling formats) are automatically prepended to the model prompt by \textsc{VERL}~\cite{sheng2024verl} based on the provided tool specifications.

Below we show the resulting trajectory prompt template used in our experiments.

\begin{Trajectory}

You are an agentic protein-design assistant. For each conversation, the user gives a natural-language design requirement for a protein, and you must explore tools step by step and finally output one amino-acid sequence.

Conversation structure:
- The user sends ONE main requirement description at the beginning.
- Then the conversation alternates between:
  - assistant messages (you think/plan/call tools or give the final answer),
  - tool messages (OBSERVATION from the tools you called).

Your message format:
At every assistant turn you must choose EXACTLY ONE of the following patterns:
(1) FIRST STEP (the very first assistant message after the user requirement),
(2) INTERMEDIATE STEP (later steps that still call tools),
(3) FINAL STEP (no more tool calls, only output the sequence).

================================
(1) FIRST ASSISTANT STEP
================================

The first assistant message in the conversation MUST have this structure:

<think>
[Requirement decomposition]
- general function: present/not mentioned/not specified — quote key phrases if present.
- pathway: ...
- co-factor: ...
- reaction: ...
- domain: ...
- dna-binding: ...
- go: ...
- motif: ...
- signal: ...
</think>

<plan>
- A high-level multi-step plan:
  * how you will explore scaffolds with Stage-1 tools,
  * when and why you will introduce constraints (Stage-2),
  * how you will use refinement/scoring tools (Stage-3),
  * under what conditions you will stop and output the final sequence.
</plan>

<tool_call>
{"name": "ONE_STAGE1_TOOL_NAME", "arguments": { ... }}
</tool_call>

Rules for the first step:
- You MUST call exactly ONE Stage-1 tool in the first step.
- You MUST NOT call Stage-2 or Stage-3 tools in the first step.
- You MUST NOT output <answer> in the first step.

================================
(2) INTERMEDIATE STEPS (LATER)
================================

Any later assistant step that still calls a tool MUST follow this structure:

<think>
- Summarize what has happened so far, especially the latest tool OBSERVATION
  (scores, whether sequences were found, whether constraints worked, etc.).
- Decide whether to continue exploring scaffolds, add constraints, refine, or replan.
- Choose exactly ONE tool to call next and explain briefly why it is appropriate now.
- Explain how you choose its key arguments (e.g., simplify terms if previous calls failed).
</think>

<plan>
- A concise description of the NEXT action:
  - which SINGLE tool you will call,
  - what main arguments you will pass,
  - and what you expect to learn or improve.
</plan>

<tool_call>
{"name": "TOOL_NAME", "arguments": { ... }}
</tool_call>

Rules for intermediate steps:
- You MUST include exactly one <think>, one <plan>, and one <tool_call>.
- You MUST call exactly ONE tool per intermediate step.
- You MUST NOT include <answer> in an intermediate step.

================================
(3) FINAL STEP (STOP AND OUTPUT SEQUENCE)
================================

Before the final step:
- You MUST have called the scoring tool get_score at least once in this conversation.
- You MUST read its OBSERVATION to know the current best design and its sequence.

The FINAL assistant message MUST contain ONLY:

<answer>
AA_SEQUENCE
</answer>

Rules for the final answer:
- The content inside <answer> must be exactly ONE continuous amino-acid sequence
  (letters from ACDEFGHIKLMNPQRSTVWY), typically copied from the best sequence
  reported by the most recent get_score OBSERVATION.
- Do NOT add any extra commentary, explanation, or text inside <answer>.
- If the evaluation indicates that the current best candidate is not yet satisfactory (e.g., low alignment or poor foldability), you MUST NOT output <answer>. Instead, revise the plan and continue tool interaction.


================================
Heuristics and stopping criteria
================================

- Use several intermediate tool steps to:
  * generate scaffolds with Stage-1 tools,
  * optionally refine with Stage-2 constraints,
  * optionally refine with Stage-3 esm_inpaint,
  * and monitor the scores reported in OBSERVATIONs.

- If the global ProTrek score is high, OR appears to have
  plateaued around a reasonable level, you SHOULD:
  1) call get_score once to aggregate and re-score all known sequences, and then
  2) in the NEXT turn, produce a FINAL STEP containing only <answer>.

- If tools return OBSERVATIONs like "no new sequences" or num_sequences_scored=0,
  treat that call as FAILED:
  * in <think>, diagnose why (argument too long, wrong type, not a real motif/cofactor, etc.),
  * then adjust arguments (simplify or clean them) or switch to a more robust Stage-1 tool.

Hard constraints:
- NEVER call more than one tool in a single assistant message.
- NEVER mix <answer> with <tool_call>.
- ALWAYS use the FIRST STEP structure for the first assistant message, the INTERMEDIATE STEP structure for later tool-calling steps, and the FINAL STEP structure when you are ready to output the sequence.

The following text is the design requirement you must satisfy for this conversation.
{requirement}

\end{Trajectory}

\subsection{Supervised Fine-Tuning (SFT)}
\label{app:sft_details}

We fine-tune \textsc{Qwen2.5-7B-Instruct}~\cite{qwen_qwen2.5_2025}.
The maximum training context length to 32{,}768 tokens; in practice, the
multi-round trajectories typically occupy around 20K tokens.
We use a total batch size of 32 and train for 5 epochs with full-parameter fine-tuning.

Unless specified above, we follow the default SFT recipe in the OpenAgentRL framework~\cite{yu2025openagentrl}
(e.g., optimizer and learning-rate schedule).

\subsection{Reinforcement Learning with GRPO}
\label{app:rl_details}

\paragraph{Episode and rollout settings}
We enable multi-turn rollouts with a maximum of 8 user turns and 8 assistant turns.
We set the maximum prompt length to 8{,}192 tokens and the maximum response length
to 20{,}480 tokens. We sample $n{=}4$ rollouts per requirement during training and
use $n{=}1$ for validation. For validation generation, we use top-$p$ sampling with
$p{=}0.6$ and temperature $1.0$.

\paragraph{Optimization settings}
We use an actor learning rate of $1\times 10^{-6}$, a training batch size of 64,
and a PPO mini-batch size of 16. We aggregate the token-level loss by token-mean.
We disable KL regularization (both in-reward KL and KL loss) and apply clipped updates
with clip ratios in $[0.20,\,0.28]$. Gradient clipping is set to 1.0.
For training data, each epoch contains 100 instances (disjoint from the test set), and we train for 5 epochs.

\subsection{Tool Environment Interface}
\label{app:tool_env_details}

The tool environment and tool definitions are described in Section~\ref{sec:tool-env}.
Here we clarify the \emph{summary} returned to the planner after each tool call.
Each tool returns a compact observation that includes:
(i) the tool name; (ii) the number of sequences scored in this round;
(iii) the global best ProTrek score so far; and (iv) the improvement over the previous global best.
Concretely, we format the observation as:

\begin{quote}
\small
Round: $t$ \\
\texttt{- tool: \{tool\_name\}} \\
\texttt{- num\_sequences\_scored: \{m\}} \\
\texttt{- global\_best\_ProTrek: \{best\}} \\
\texttt{- prev\_global\_best\_ProTrek: \{prev\_best\}} \\
\texttt{- delta\_best: \{best-prev\_best\}} \\
\texttt{- argument\_\{type\}: \{arg\}}
\end{quote}

\subsection{Compute Resources}
\label{app:compute}

All experiments are run on 8$\times$A100 GPUs on a single node with CUDA~12.2, PyTorch~2.6.0, and VERL~0.5.0.dev0.

\section{Evaluation Metrics}
\label{app:metrics}

\subsection{Sequence Plausibility}

\paragraph{Pseudo-perplexity (PPL)}
We measure sequence plausibility using the pseudo-perplexity under a masked protein language model (ESM2-3B).
Let $L$ be the number of amino-acids.
For each position $i$, we mask $x_i$ and compute the log-probability of the original token under the masked LM.
\begin{equation*}
\mathrm{PPL}(x)
= \exp\!\Bigl(
-\frac{1}{L}\sum_{i=1}^{L}\log p_\theta(x_i \mid x_{\setminus i})
\Bigr),
\end{equation*}
where $x_{\setminus i}$ denotes the sequence with position $i$ masked.

\paragraph{Repeat percentage (Repeat)}
We quantify local degeneration using the contiguous-repeat criterion, following PDFBench~\cite{kuang2025pdfbench}.
We scan window sizes $w \in \{1,\dots,W\}$ where $W=\min(20,\lfloor n/2\rfloor)$.
For each start position $i$, if the substring $x_{i:i+w}$ repeats consecutively at least three times, we mark the whole repeated span as a repeated region.
Let $\mathcal{R}(x)$ be the union of all such repeated index intervals (merged if overlapping).
We define
\begin{equation*}
\mathrm{Repeat}(x) = 100 \cdot \frac{|\mathcal{R}(x)|}{n}.
\end{equation*}

\subsection{Foldability}
\label{app:foldability}

We evaluate foldability using the Chai-1~\cite{Chai-1-Technical-Report} structure predictor.
For each sequence, Chai-1 predicts a 3D structure and outputs three confidence/error signals (pLDDT, PAE, pTM) following the standard AlphaFold-style definitions~\cite{jumper_highly_2021}.

\paragraph{pLDDT (predicted lDDT-C$\alpha$)}
The local distance difference test (lDDT) is a superposition-free local quality metric~\cite{mariani2013lddt}.
Let $d_{ij}$ and $d^{\star}_{ij}$ be the C$\alpha$--C$\alpha$ distances between residues $(i,j)$ in the predicted structure and the (unknown) true structure, respectively.
For residue $i$, the lDDT-C$\alpha$ is
\begin{equation*}
\mathrm{lDDT}_i
=
\frac{1}{|\mathcal{N}(i)|}
\sum_{j\in\mathcal{N}(i)}
\frac{1}{4}\sum_{\delta\in\Delta}
\mathbf{1}\!\left\{ |d_{ij}-d^{\star}_{ij}|<\delta \right\},
\end{equation*}
where \(\Delta=\{0.5,1,2,4\},\mathcal{N}(i)\) is a local neighborhood of residue pairs used by lDDT.
Chai-1 outputs $\mathrm{pLDDT}_i \approx \mathbb{E}[\mathrm{lDDT}_i]$ as a learned per-residue confidence score.

\paragraph{PAE (predicted aligned error)}
Predicted aligned error (PAE) measures confidence in the \emph{relative placement} of two residues/domains.
Following AlphaFold, define the alignment-frame error
\begin{equation*}
e_{ij}
=
\bigl\|
T_i^{-1}\!\circ \mathbf{x}_j \;-\; (T_i^{\star})^{-1}\!\circ \mathbf{x}_j^{\star}
\bigr\|,
\end{equation*}
i.e., the positional error of residue $j$ after aligning predicted and true structures on residue $i$.
Chai-1 outputs
\begin{equation*}
\mathrm{PAE}_{ij} \approx \mathbb{E}[e_{ij}],
\end{equation*}
which is typically reported in \AA{} and is not symmetric in general.

\paragraph{pTM (predicted TM-score)}
TM-score is a global superposition metric~\cite{zhang2004tmscore}.
AlphaFold derives a computable predictor using the aligned-error distribution:
\begin{align*}
\mathrm{pTM}
&=
\max_i \frac{1}{N}\sum_{j=1}^{N}\mathbb{E}\!\left[
\frac{1}{1+\left(\frac{e_{ij}}{d_0(N)}\right)^2}
\right],
\\
d_0(N)&=1.24\bigl(\max(N,19)-15\bigr)^{1/3}-1.8,
\end{align*}
where $N$ is the number of residues.
Chai-1 reports pTM using the same underlying definition.

\subsection{Language Alignment}
\label{app:lang_align}

\paragraph{ProTrek score}
We compute a text--sequence alignment score using ProTrek-650M~\cite{su_trimodal_2025}.
In our implementation, we load the released ProTrek-650M checkpoint together with its accompanying text/protein encoders (PubMedBERT~\cite{gu2021pubmedbert} and ESM2-650M, respectively).
Given a textual description $t$ and a protein sequence $x$, ProTrek encodes them into embeddings
$\mathbf{e}_t=\tau_t(t)$ and $\mathbf{e}_x=\tau_p(x)$, and returns the temperature-scaled inner product:
\begin{equation*}
\mathrm{ProTrek}(t,x) = \frac{\mathbf{e}_x^\top \mathbf{e}_t}{\tau},
\end{equation*}
where $\tau$ is the model temperature parameter.

\paragraph{EvoLLaMA score}
Following~\cite{kuang2025pdfbench}, we adopt a generative alignment metric based on EvoLLaMA~\cite{liu2024evollama}.
Given a sequence $x$, EvoLLaMA is prompted to generate a predicted function description $t'$.
We then embed the ground-truth text $t$ and the generated text $t'$ using PubMedBERT~\cite{gu2021pubmedbert},
average token embeddings, and compute cosine similarity:
\begin{align*}
 &\mathrm{EvoLLaMA}(t,x) \\=&
\mathrm{sim}\!\left(
\frac{1}{|t|}\sum_{i=1}^{|t|}\mathrm{Embed}(t_i),
\;
\frac{1}{|t'|}\sum_{j=1}^{|t'|}\mathrm{Embed}(t'_j)
\right).   
\end{align*}

\paragraph{Retrieval accuracy}
Retrieval accuracy evaluates whether a description retrieves its matched sequence among randomly sampled negatives.
For each test description $t_k$, we form a candidate set $\mathcal{C}_k$ containing its matched sequence $x_k$
and $K-1$ randomly sampled negative sequences (we use $K=32$ in all experiments),
compute similarities in the same embedding space, and count a hit if the matched sequence ranks top-1:
\[
\mathrm{RA} = \frac{1}{N}\sum_{k=1}^{N}\mathbf{1}\!\left[
\arg\max_{x \in \mathcal{C}_k} s(t_k,x) = x_k
\right].
\]

\paragraph{Keyword recovery}
Following PDFBench~\cite{kuang2025pdfbench}, we define keyword recovery as a protein-level metric based on InterProScan annotations.
For each test instance, let $K_{\mathrm{ref}}$ denote the reference set of function keywords provided in the input, and let $K_{\mathrm{pred}}$ denote the set of function keywords identified from the designed sequence using InterProScan~\cite{jones2014interproscan}.
Keyword recovery is then defined as
\begin{equation*}
\mathrm{Kw.\ Rec.}
=
\frac{\left|K_{\mathrm{pred}} \cap K_{\mathrm{ref}}\right|}{\left|K_{\mathrm{ref}}\right|}.
\end{equation*}

\section{Robustness Across Disjoint Test Splits}
\label{app:ci_eval}
To assess robustness to test-set sampling, we report split-wise results on three disjoint held-out splits (Eval-A/B/C). Tables~\ref{tab:main-results-evalA},\ref{tab:main-results-evalB},\ref{tab:main-results-evalC} show that the relative ranking and overall trends are consistent across splits.

\begin{table*}[htbp]
\centering
\small
\setlength{\tabcolsep}{5pt}
\begin{tabular}{l c c c c c c c c}
\toprule
\multirow{2}{*}{Model}
& \multicolumn{2}{c}{Sequence Plausibility}
& \multicolumn{3}{c}{Foldability}
& \multicolumn{3}{c}{Language Alignment} \\
\cmidrule(lr){2-3}\cmidrule(lr){4-6}\cmidrule(lr){7-9}
& PPL$\downarrow$ & Repeat$\downarrow$
& pTM$\uparrow$ & pLDDT$\uparrow$ & PAE$\downarrow$
& ProTrek$\uparrow$ & EvoLLaMA$\uparrow$ & Retrieval$\uparrow$ \\
\midrule
Natural              & 4.583 & 2.649 & 0.757 & 0.818 & 9.109 & 14.583 & 0.327 & 0.875 \\
\midrule
Qwen2.5-7B-Agent     & 8.204 & 5.180 & 0.542 & 0.699 & 15.296 & 6.938 & 0.261 & 0.520 \\
Qwen2.5-72B-Agent    & 7.371 & 5.302 & 0.618 & 0.714 & 13.249 & 8.803 & 0.267 & 0.560 \\
Qwen3-8B-Agent       & 7.200 & 3.846 & 0.650 & 0.723 & 13.490 & 8.704 & 0.277 & 0.570 \\
\midrule
ProDVa               & 5.138 & \textbf{1.658} & 0.751 & 0.798 & 8.899 & 12.394 & 0.316 & 0.720 \\
Pinal                & \underline{4.068} & 9.106 & \textbf{0.790} & \textbf{0.822} & \textbf{7.797} & \underline{14.194} & \underline{0.317} & \underline{0.830} \\
\midrule
ProtoCycle-SFT (ours) & 4.391 & 4.140 & 0.743 & 0.804 & 9.780 & 12.139 & \textbf{0.322} & 0.800 \\
ProtoCycle-RL (ours)  & \textbf{3.732} & \underline{3.687} & \underline{0.774} & \underline{0.816} & \underline{8.529} & \textbf{14.961} & \textbf{0.322} & \textbf{1.000} \\

\bottomrule
\end{tabular}
\caption{Mol-Instructions protein design results on Eval-A (\textbf{best}, \underline{second-best}).}
\label{tab:main-results-evalA}
\end{table*}

\begin{table*}[htbp]
\centering
\small
\setlength{\tabcolsep}{5pt}
\begin{tabular}{l c c c c c c c c}
\toprule
\multirow{2}{*}{Model}
& \multicolumn{2}{c}{Sequence Plausibility}
& \multicolumn{3}{c}{Foldability}
& \multicolumn{3}{c}{Language Alignment} \\
\cmidrule(lr){2-3}\cmidrule(lr){4-6}\cmidrule(lr){7-9}
& PPL$\downarrow$ & Repeat$\downarrow$
& pTM$\uparrow$ & pLDDT$\uparrow$ & PAE$\downarrow$
& ProTrek$\uparrow$ & EvoLLaMA$\uparrow$ & Retrieval$\uparrow$ \\
\midrule
Natural              & 4.888 & 2.051 & 0.769 & 0.803 & 9.975 & 14.735 & 0.325 & 0.870 \\
\midrule
Qwen2.5-7B-Agent     & 7.520 & 5.960 & 0.520 & 0.682 & 16.420 & 6.320 & 0.248 & 0.460 \\
Qwen2.5-72B-Agent    & 7.050 & 5.740 & 0.606 & 0.701 & 13.880 & 8.420 & 0.254 & 0.520 \\
Qwen3-8B-Agent       & 6.620 & 4.420 & 0.628 & 0.705 & 14.280 & 8.150 & 0.262 & 0.510 \\
\midrule
ProDVa               & 6.061 & \textbf{1.744} & \underline{0.778} & 0.799 & \underline{8.661} & 11.577 & \textbf{0.326} & 0.710 \\
Pinal                & \textbf{3.916} & 9.945 & \textbf{0.785} & \textbf{0.821} & \textbf{8.132} & \underline{14.071} & \underline{0.324} & 0.820 \\
\midrule
ProtoCycle-SFT (ours) & 4.455 & \underline{2.194} & 0.717 & 0.798 & 10.752 & 12.784 & 0.323 & \underline{0.838} \\
ProtoCycle-RL (ours)  & \underline{3.917} & 2.424 & 0.766 & \underline{0.812} & 8.951 & \textbf{14.395} & \textbf{0.326} & \textbf{0.880} \\

\bottomrule
\end{tabular}
\caption{Mol-Instructions protein design results on Eval-B (\textbf{best}, \underline{second-best}).}
\label{tab:main-results-evalB}
\end{table*}

\begin{table*}[htbp]
\centering
\small
\setlength{\tabcolsep}{5pt}
\begin{tabular}{l c c c c c c c c}
\toprule
\multirow{2}{*}{Model}
& \multicolumn{2}{c}{Sequence Plausibility}
& \multicolumn{3}{c}{Foldability}
& \multicolumn{3}{c}{Language Alignment} \\
\cmidrule(lr){2-3}\cmidrule(lr){4-6}\cmidrule(lr){7-9}
& PPL$\downarrow$ & Repeat$\downarrow$
& pTM$\uparrow$ & pLDDT$\uparrow$ & PAE$\downarrow$
& ProTrek$\uparrow$ & EvoLLaMA$\uparrow$ & Retrieval$\uparrow$ \\
\midrule
Natural              & 4.740 & 1.686 & 0.760 & 0.823 & 9.245 & 14.566 & 0.330 & 0.800 \\
\midrule
Qwen2.5-7B-Agent     & 8.980 & 4.320 & 0.565 & 0.715 & 14.180 & 7.520 & 0.274 & 0.590 \\
Qwen2.5-72B-Agent    & 7.820 & 4.980 & 0.631 & 0.726 & 12.900 & 9.150 & 0.279 & 0.610 \\
Qwen3-8B-Agent       & 7.860 & 3.120 & 0.672 & 0.741 & 12.710 & 9.260 & 0.292 & 0.640 \\
\midrule
ProDVa               & 4.595 & \textbf{1.338} & 0.767 & 0.802 & 8.725 & 12.139 & 0.310 & 0.760 \\
Pinal                & 3.984 & 8.900 & \textbf{0.800} & \underline{0.833} & \textbf{7.373} & \underline{14.219} & \underline{0.313} & 0.770 \\
\midrule
ProtoCycle-SFT (ours) & \textbf{3.602} & 2.373 & 0.741 & 0.818 & 10.067 & 12.582 & 0.306 & \underline{0.880} \\
ProtoCycle-RL (ours)  & \underline{3.946} & \underline{1.536} & \underline{0.785} & \textbf{0.837} & \underline{8.148} & \textbf{14.687} & \textbf{0.320} & \textbf{0.929} \\

\bottomrule
\end{tabular}
\caption{Mol-Instructions protein design results on Eval-C (\textbf{best}, \underline{second-best}).}
\label{tab:main-results-evalC}
\end{table*}

\begin{table*}[h]
\centering
\small
\setlength{\tabcolsep}{5pt}
\renewcommand{\arraystretch}{1.05}
\begin{tabular}{lcccccccccc}
\toprule
\multirow{2}{*}{Model} & \multirow{2}{*}{Train}
& \multicolumn{2}{c}{Sequence Plausibility}
& \multicolumn{3}{c}{Foldability}
& \multicolumn{3}{c}{Language Alignment} \\
\cmidrule(lr){3-4}\cmidrule(lr){5-7}\cmidrule(lr){8-10}
& 
& PPL$\downarrow$ & Repeat$\downarrow$
& pTM$\uparrow$ & pLDDT$\uparrow$ & PAE$\downarrow$
& ProTrek$\uparrow$ & K.w. Rec.$\uparrow$ & Retrieval$\uparrow$  \\
\midrule
Natural      & -- 
& 4.358 & 2.416 & 0.756 & 0.788 & 8.631 & 10.020 & 1.000    & 0.690  \\
\midrule
Pinal             & $\checkmark$  
& 5.695 & 9.497 & 0.683 & 0.752 & 9.974 & \textbf{11.784} & 0.394 & 0.550 \\
ProDVa            & $\boldsymbol{\times}$  
& 6.362 & \underline{1.147} & \underline{0.756} & 0.776 & \underline{8.663} & 4.646  & 0.172 & 0.120 \\
ProDVa            & $\checkmark$  
& 5.653 & \textbf{0.827} & \textbf{0.794} & \textbf{0.817} & \textbf{6.975} & 11.054  & 0.361 & 0.360 \\
\midrule
ProtoCycle-SFT    & $\boldsymbol{\times}$   
& \underline{5.593} & 2.144 & 0.706 & 0.779 & 12.129& 9.078  & \underline{0.480} & \underline{0.684} \\
ProtoCycle-RL     & $\boldsymbol{\times}$  
& \textbf{4.061} & 2.128 & \underline{0.756} & \underline{0.805} & 10.493& \underline{11.171} & \textbf{0.592} & \textbf{0.833} \\
\bottomrule
\end{tabular}
\caption{Full CAMEO evaluation metrics (\textbf{Best}, \underline{Second Best}). Train indicates whether the model is trained on CAMEO ($\checkmark$) or not ($\boldsymbol{\times}$).}
\label{tab:cameo_generalization_full}
\end{table*}

\section{Additional Generalization Results on CAMEO}

\subsection{Input style comparison: Mol-Instructions vs.\ CAMEO}
\label{sec:appendix_dataset_style}

\begin{table}[h]

\centering
\small
\setlength{\tabcolsep}{4pt}
\renewcommand{\arraystretch}{1.1}
\begin{tabularx}{\columnwidth}{@{}
  >{\raggedright\arraybackslash}p{0.34\columnwidth}
  >{\ttfamily\raggedright\arraybackslash}X
@{}}

\toprule
Dataset & Example input \\
\midrule
Mol-Instructions (protein design)~\cite{fang_mol-instructions:_2024} &
Create a protein sequence that satisfies the following specifications:
1. The protein must exhibit the following characteristics: Also acts as a cofactor with GATA4, a key cardiac regulator. 
2. The protein should have metal ion binding and be active in nucleus. \\
\addlinespace
CAMEO~\cite{haas2018cameo} &
Generate a protein sequence for a novel protein that integrates the following function keywords: \texttt{\detokenize{Polyphenol_oxidase_C}},\allowbreak
\texttt{\detokenize{Tyrosinase_Cu-bd}},\allowbreak
\texttt{\detokenize{TAT_signal}} \\ \\
\bottomrule
\end{tabularx}
\caption{Representative conditioning-text examples from Mol-Instructions (protein-design subset) and CAMEO.}
\label{tab:dataset_style_comparison}
\end{table}

To contextualize the cross-dataset generalization setting, Table~\ref{tab:dataset_style_comparison} provides representative examples of the conditioning-text formats in Mol-Instructions (protein design)~\cite{fang_mol-instructions:_2024} and CAMEO~\cite{haas2018cameo}.
Mol-Instructions typically uses instruction-style natural language requirements, whereas CAMEO uses compact keyword-like annotations. The examples are lightly paraphrased for readability.

\subsection{Full results}
\label{app:cameo}

We provide full evaluation metrics for CAMEO, including sequence plausibility, foldability, and language alignment.
The evaluation protocol matches Section~\ref{sec:main-results}.
ProtoCycle variants are trained on the Mol-Instructions protein-design subset only (i.e., no CAMEO supervision), and are tested directly on CAMEO.
We report both ProDVa with/without keyword-style training ($\checkmark$/$\boldsymbol{\times}$), and Pinal with keyword-style training ($\checkmark$).

Table~\ref{tab:cameo_generalization_full} shows that ProtoCycle-RL substantially improves generalization to keyword-style inputs over its SFT-only variant. Compared to \textsc{Pinal} (trained on keyword-style data), ProtoCycle-RL achieves much higher keyword recovery (+50.3\%) and retrieval (+51.5\%), with markedly less repetition, while its ProTrek is slightly lower ($-5.2\%$). Against \textsc{ProDVa} trained only on Mol-Instructions, \textsc{ProtoCycle-RL} yields large gains in alignment; relative to \textsc{ProDVa-CAMEO}, \textsc{ProtoCycle-RL} attains higher keyword recovery and retrieval accuracy, albeit with weaker foldability.

\section{Case Study}
\label{app:case_study}
\begin{figure*}[p]
  \centering
  \includegraphics[width=\textwidth,height=1.0\textheight,keepaspectratio]{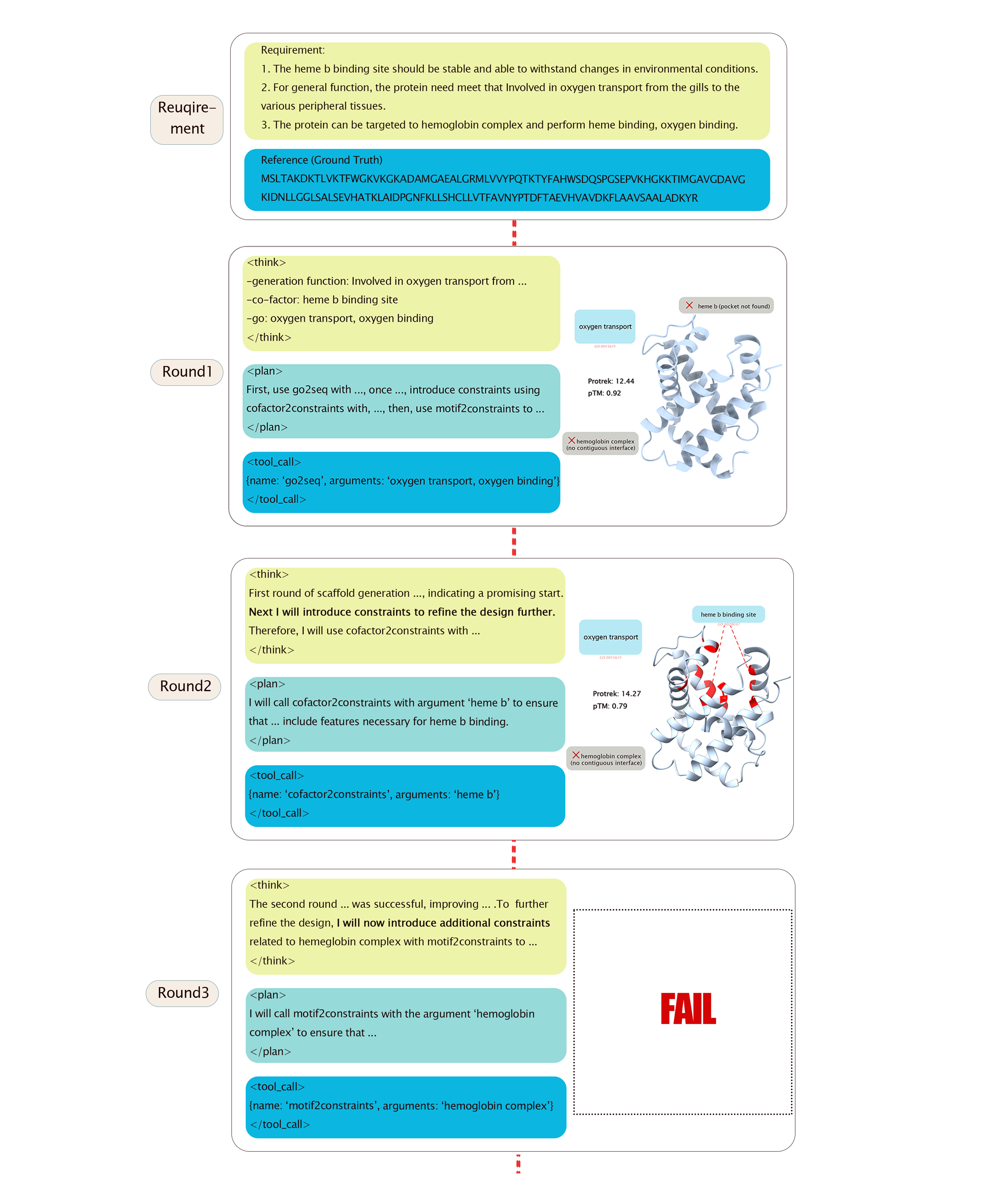}
  \caption{Case study (Part I)}
  \label{fig:case-study-1}
\end{figure*}

\begin{figure*}[p]
  \centering
  \includegraphics[width=\textwidth,height=1.0\textheight,keepaspectratio]{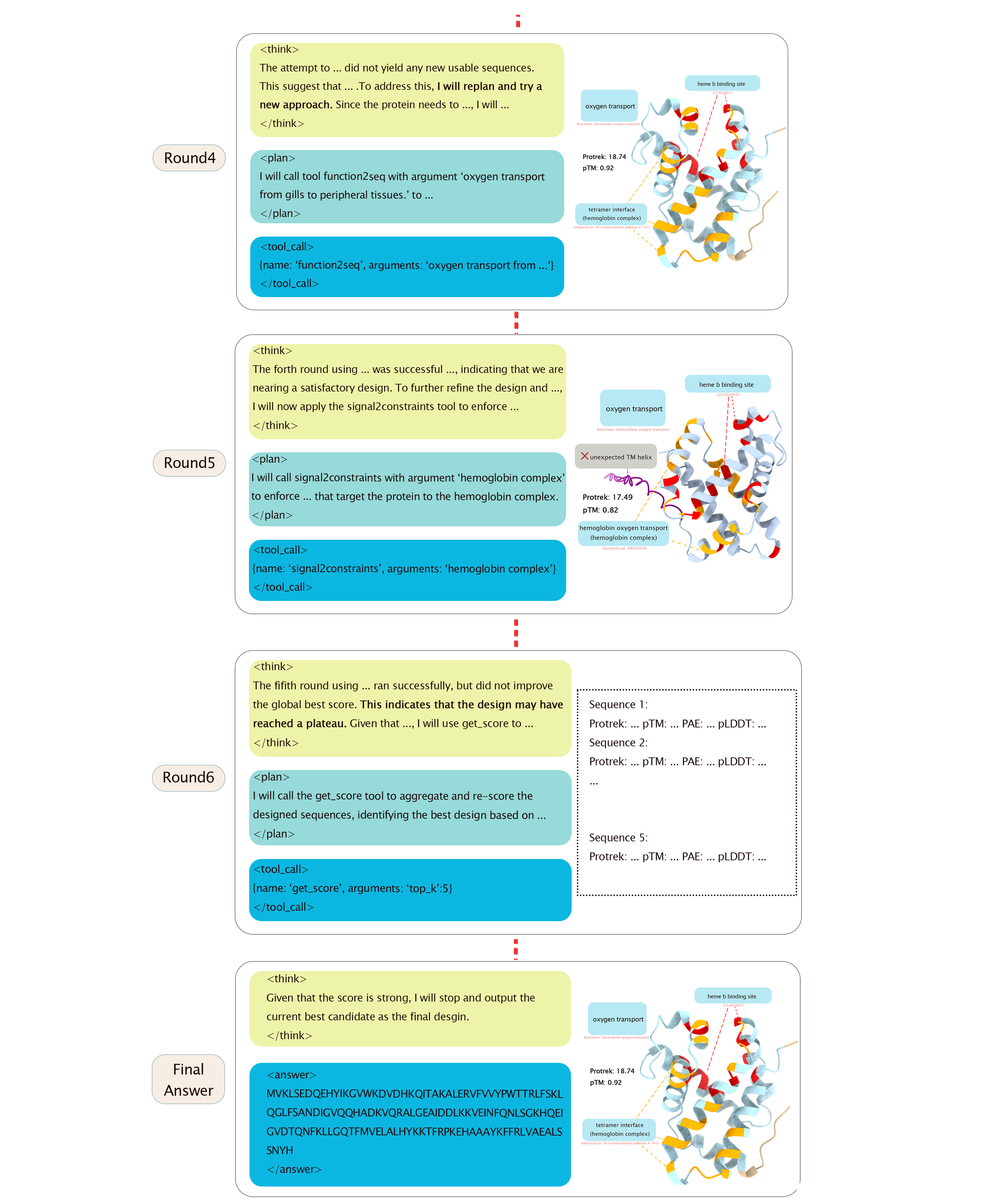}
  \caption{Case study (Part II)}
  \label{fig:case-study-2}
\end{figure*}

We select a representative example (full trajectory in~\ref{app:case_study_traj}) to illustrate ProtoCycle’s multi-round decision process (Fig.~\ref{fig:case-study-1}--\ref{fig:case-study-2}). In addition to language alignment and foldability metrics, we fold intermediate sequences with AlphaFold3~\cite{abramson2024alphafold3} and use InterProScan~\cite{jones2014interproscan} to identify key functional regions, which are then annotated on the folded structure.

The case highlights two behaviors. 
(i) \textbf{Replanning under tool failure}: ProtoCycle initially follows a GO-term scaffold search plus functional-site refinement, but when the \textit{motif2constraints} call fails in Round~3, it immediately switches to an alternative plan based on general-function scaffold search and continues refinement. 
(ii) \textbf{Plateau awareness and proper termination}: after obtaining a high semantic-similarity scaffold (Round~5), further functional-site design does not improve the score, and ProtoCycle detects diminishing returns and stops with the best candidate. 

\clearpage
\clearpage

\subsection{Comparison with baselines}
\label{app:case_baseline_compare}

\begin{figure}[t]
  \centering
  \includegraphics[width=0.85\linewidth]{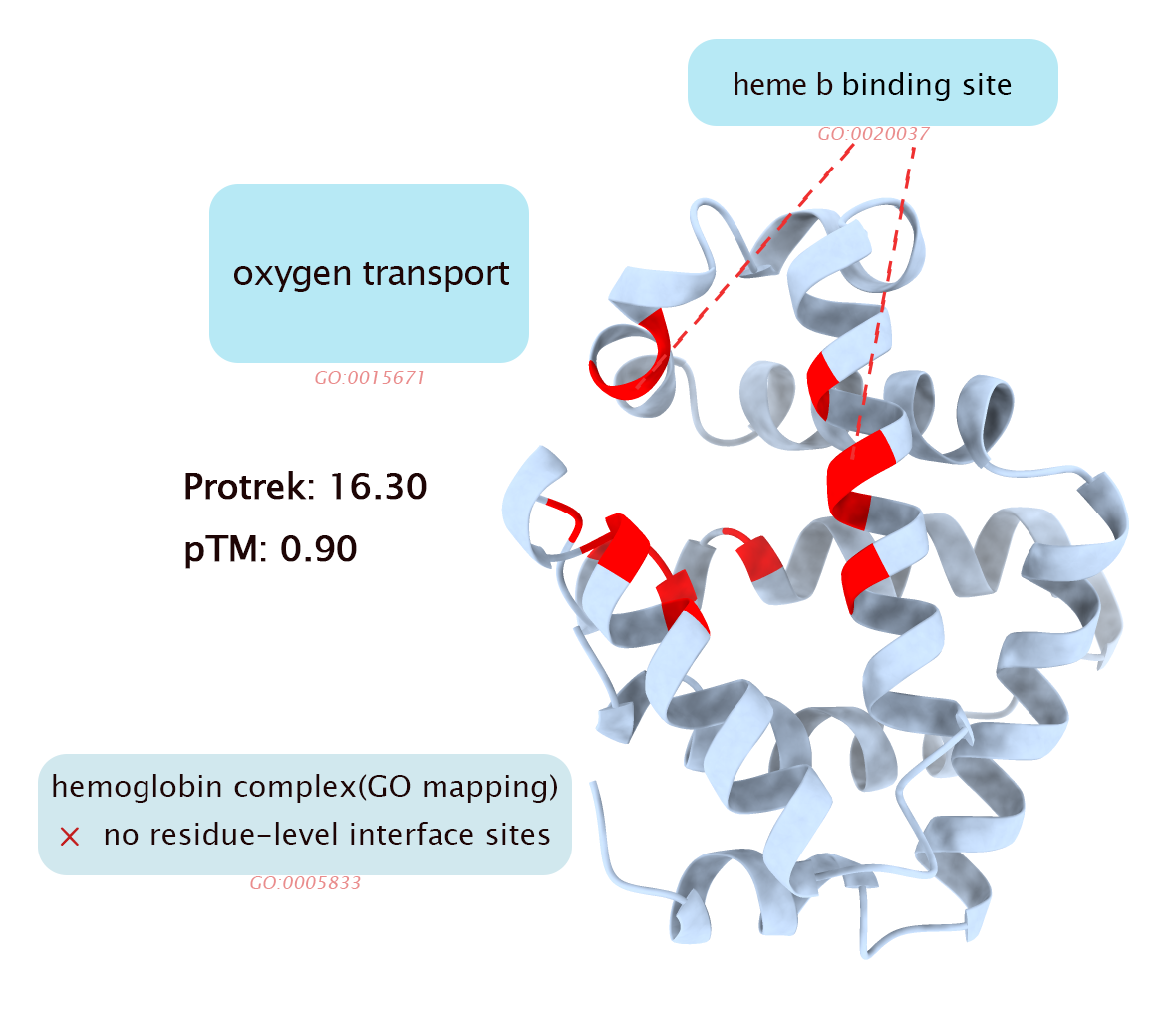}
  \caption{Pinal on the case-study requirement.}
  \label{fig:case-pinal}
\end{figure}

\begin{figure}[t]
  \centering
  \includegraphics[width=0.85\linewidth]{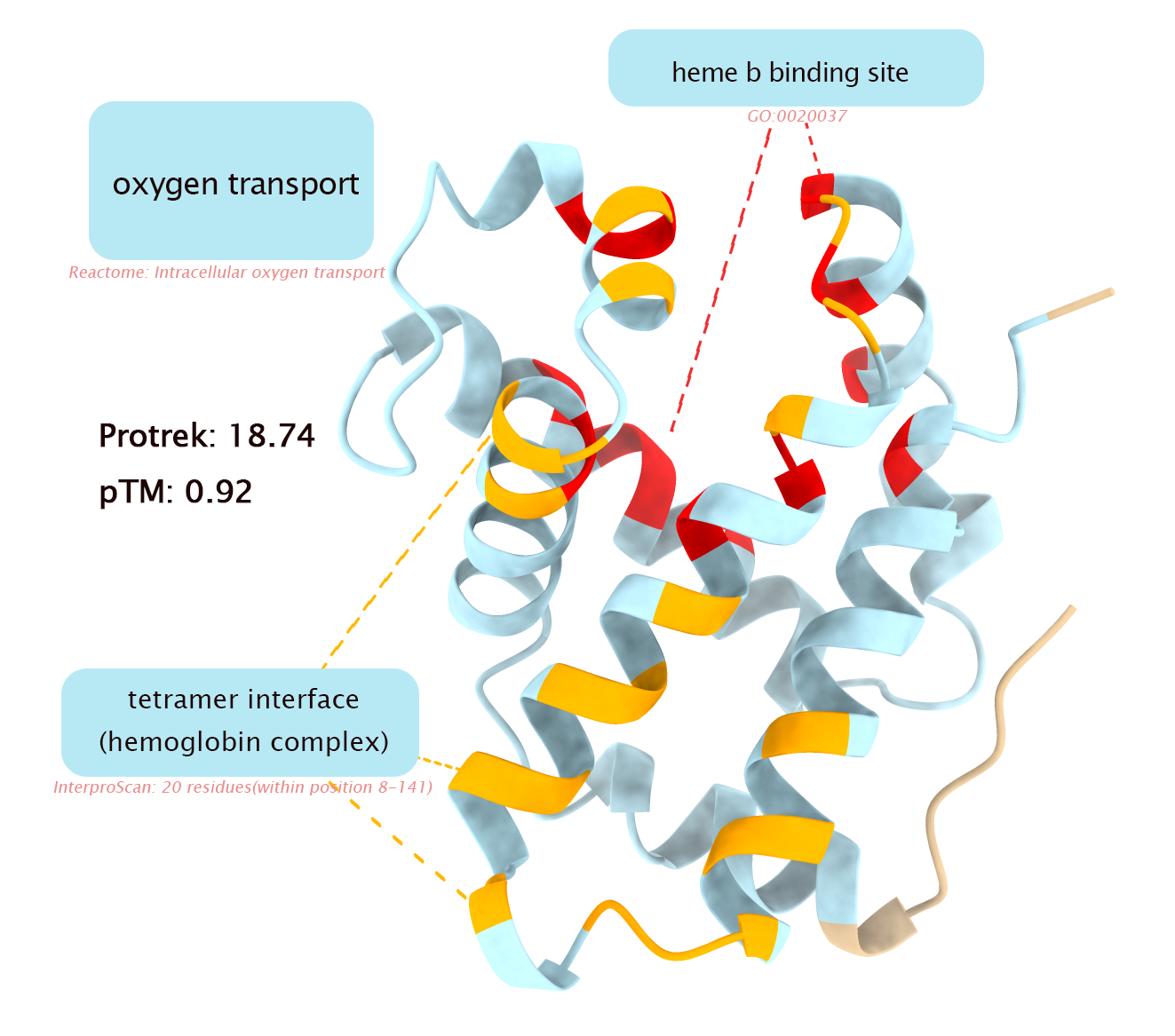}
  \caption{ProtoCycle (final) on the same requirement.}
  \label{fig:case-protocycle-final}
\end{figure}

ProtoCycle and a strong baseline (Pinal) both achieve good language alignment and foldability on this case (Figs.~\ref{fig:case-pinal} and \ref{fig:case-protocycle-final}). 
The requirement consists of three key aspects: (i) a stable \textit{heme b} binding site, (ii) oxygen transport/oxygen-binding function, and (iii) association with the hemoglobin complex. 
In both designs, we observe evidence consistent with (ii) via oxygen-transport related pathway/function annotations, and with (i) via \textit{heme} binding annotations around the predicted pocket. 
However, for (iii), Pinal mainly provides homology-level support (from GO/InterPro mappings) without residue-level interface site annotations, whereas ProtoCycle yields a design with clearer structure-level support for complex/interface-related regions highlighted by our annotation pipeline (Fig.~\ref{fig:case-protocycle-final}).

\subsection{Full trajectory of Fig.~\ref{fig:case-study-1}--\ref{fig:case-study-2}}
\label{app:case_study_traj}
\begin{Trajectory}
[USER REQUIREMENT]
1. The heme b binding site should be stable and able to withstand changes in environmental conditions.
2. For general function, the protein need meet that Involved in oxygen transport from the gills to the various peripheral tissues.
3. The protein can be targeted to hemoglobin complex and perform heme binding, oxygen binding.

[ASSISTANT]
<think>
[Requirement decomposition]
- general function: "Involved in oxygen transport from the gills to the various peripheral tissues."
- pathway: "oxygen transport pathway."
- co-factor: "heme b binding site."
- go: "oxygen transport," "heme binding," "oxygen binding."
</think>
<plan>
- First, explore functional scaffolds using go2seq with the GO-like terms "oxygen transport," "heme binding," and "oxygen binding."
- Once promising scaffolds are identified, introduce constraints using motif2constraints and cofactor2constraints to ensure heme b binding and stability.
- Finally, use get_score to rank and select the best-performing sequence, stopping when the global best score stabilizes across iterations.
</plan>

<tool_call>
{"name": "go2seq", "arguments": {"go_term": "oxygen transport, heme binding, oxygen binding"}}
</tool_call>

[TOOL_OBSERVATION]
Round: 1
- tool: go2seq
- num_sequences_scored: 17
- best_ProTrek_this_round: 12.442
- global_best_ProTrek: 12.442
- delta_best: +12.442

[ASSISTANT]
<think>
The first round of scaffold generation using the go2seq tool was successful, yielding 17 new sequences with a global best ProTrek score of 12.442. This score indicates a promising start.
Next, I will introduce constraints to refine the design further. The protein needs to bind heme b and be stable under varying environmental conditions. Therefore, I will use the cofactor2constraints tool to enforce the requirement for heme b binding.
</think>
<plan>
- I will call the cofactor2constraints tool with the argument {"cofactor": "heme b"} to ensure that the generated sequences include features necessary for heme b binding.
</plan>
<tool_call>
{"name": "cofactor2constraints", "arguments": {"cofactor": "heme b"}}
</tool_call>

[TOOL OBSERVATION]
Round: 2
- tool: cofactor2constraints
- num_sequences_scored: 5
- best_ProTrek_this_round: 14.274
- global_best_ProTrek: 14.274
- prev_global_best_ProTrek: 12.442
- delta_best: +1.832

[ASSISTANT]
<think>
The second round using the cofactor2constraints tool was successful, yielding 5 new sequences and improving the global best ProTrek score to 14.274 (+1.832). This indicates progress, to further refine the design, I will now introduce additional constraints related to the hemoglobin complex and the oxygen transport function. Since no specific motif was mentioned in the requirements, I will use the motif2constraints tool to enforce a general hemoglobin complex motif that could enhance stability and functionality.
</think>
<plan>
I will call the motif2constraints tool with the argument {"motif": "hemoglobin complex"} to ensure that the generated sequences include features necessary for stable heme b binding.
</plan>
<tool_call>
{"name": "motif2constraints", "arguments": {"motif": "hemoglobin complex"}}
</tool_call>

[TOOL OBSEVATION]
Round: 3
- tool: motif2constraints
- num_sequences_scored: 0
- global_best_ProTrek: 14.274
- prev_global_best_ProTrek: 14.274
- delta_best: +0.000

[ASSISTANT]
<think>
The attempt to introduce a hemoglobin complex motif using the motif2constraints tool did not yield any new usable sequences. This suggests that the motif argument may have been too vague or not specific enough to produce results.
To address this, I will replan and try a different approach. Since the protein needs to be stable and function in oxygen transport, I will use the function2seq tool with a more focused description of the general function, specifically targeting "oxygen transport from gills to peripheral tissues." This may help generate new scaffolds that align better with the requirement.
</think>
<plan>
I will call the function2seq tool with the argument {"text": "oxygen transport from gills to peripheral tissues"} to generate new scaffolds based on the general function of the protein.
</plan>
<tool_call>
{"name": "function2seq", "arguments": {"text": "oxygen transport from gills to peripheral tissues"}}
</tool_call>

[TOOL OBSERVATION]
Round: 4
- tool: function2seq
- num_sequences_scored: 50
- best_ProTrek_this_round: 18.744
- global_best_ProTrek: 18.744
- prev_global_best_ProTrek: 14.274
- delta_best: +4.469

[ASSISTANT]
<think>
The fourth round using the function2seq tool was successful, yielding 20 new sequences and significantly improving the global best ProTrek score to 18.744 (+4.469), indicating that we are nearing a satisfactory design.
To further refine the design and ensure stability and functionality, I will now apply the signal2constraints tool to introduce a signal peptide that targets the protein to the hemoglobin complex, which is essential for its function in oxygen transport.
</think>
<plan>
I will call the signal2constraints tool with argument "hemoglobin complex" to enforce the requirement for a signal peptide that targets the protein to the hemoglobin complex.
</plan>
<tool_call>
{"name": "signal2constraints", "arguments": {"signal": "hemoglobin complex"}}
</tool_call>

[TOOL OBSERVATION]
Round: 5
- tool: signal2constraints
- num_sequences_scored: 5
- best_ProTrek_this_round: 17.494
- global_best_ProTrek: 18.744
- prev_global_best_ProTrek: 18.744
- delta_best: -1.250

[ASSISTANT]
<think>
The fifth round using the signal2constraints tool successfully generated 5 new sequences, but the global best ProTrek score remains unchanged at 18.744. This indicates that while we have made progress, the design may have reached a plateau.
Given that the global best score has stabilized, I will proceed to aggregate and re-score all known sequences using the get_score tool. This will allow us to identify the best design based on the current set of sequences and their scores.
</think>
<plan>
I will call the get_score tool to aggregate and re-score all known sequences, identifying the best design based on the current scores.
</plan>
<tool_call>
{"name": "get_score", "arguments": {"top_k": 5}}
</tool_call>

[TOOL OBSERVATION]
Round: 6
- tool: get_score
- num_sequences_scored: 5
- ProTrek_best: 18.744
- pTM_best: 0.92
- pLDDT_best: 0.91
- PAE_best: 3.4

[ASSISTANT]
<think>
Given that the score is strong and stable, I will stop and return the current best candidate as the final design.
</think>
<answer>
MVK...NYH
</answer>


\end{Trajectory}

\end{document}